\documentclass[12pt,preprint]{aastex} % use larger type; default would be 10pt

\usepackage{graphicx} % support the \includegraphics command and options 
\usepackage{natbib} %needed for the bibliography? 
\usepackage{color} % needed to use \textcolor{color}{text}
\usepackage{float} % can use [H] to force float position 
\usepackage{amsmath} % contains math functions
\usepackage{epstopdf} % include eps figures
\usepackage{lscape} % landscape a page

\begin{document}

\hyphenpenalty=1000
 
% definitions 
\def\d2o{\hbox{$\rm{D_2O}$}} 
\def\h2o{\hbox{$\rm{H_2O}$}}
\def\wavenum{\hbox{$\rm{cm^{-1}}$}} 
\def\microns{\hbox{$\mu$m}}

\def\deg{\hbox{$^{\circ}$}}
\def\arcmin{\hbox{$^{\prime}$}}
\def\arcsec{\hbox{$^{\prime\prime}$}}

\def\hour{\hbox{$^{\rm h}$}}
\def\min{\hbox{$^{\rm m}$}}
\def\sec{\hbox{$^{\rm s}$}}

%%%%%%%%%%%%%%%%%%%%%%%%%%%%%%%%%%%%%%%%%%%%%%%%%%%%%%%%%%%%%%%%%%%%%%%%%%%%

\title{Laboratory Determination of the Infrared Band Strengths of Pyrene Frozen in Water Ice: Implications for the Composition of Interstellar Ices}

\author{E. E. Hardegree-Ullman\altaffilmark{1,2}, M. S. Gudipati\altaffilmark{3,4}, 
A. C. A. Boogert\altaffilmark{2,5}, H. Lignell\altaffilmark{6}, L. J. Allamandola\altaffilmark{7},
K. R. Stapelfeldt\altaffilmark{8}, and M. Werner\altaffilmark{3}}

\altaffiltext{1}{New York Center for Astrobiology and Department of Physics, Applied Physics, and Astronomy, Rensselaer Polytechnic Institute, 110 8$^{\rm{th}}$ Street, Troy, NY 12180, USA; hardee@rpi.edu} 
\altaffiltext{2}{Infrared Processing and Analysis Center, Mail Code 100-22, California Institute of Technology, Pasadena, CA 91125, USA}
\altaffiltext{3}{Jet Propulsion Laboratory, California Institute of Technology, 4800 Oak Grove Drive, Pasadena, CA 91109, USA}
\altaffiltext{4}{IPST, University of Maryland, College Park, MD 20742, USA; gudipati@jpl.nasa.gov}
\altaffiltext{5}{SOFIA Science Center, USRA, NASA Ames Research Center, M.S. N232-12, Moffett Field, CA 94035, USA} 
\altaffiltext{6}{Division of Chemistry and Chemical Engineering, California Institute of Technology, Pasadena, CA 91125, USA}%Department of Chemistry, University of California Irvine, Irvine, CA 92697-2025, USA}
\altaffiltext{7}{Space Science Division, Mail Stop 245-6, NASA Ames Research Center, Moffett Field, CA 94035, USA} 
\altaffiltext{8}{NASA Goddard Space Flight Center, Exoplanets and Stellar Astrophysics Laboratory, Code 667, Greenbelt, MD 20771, USA} 
%%%%%%%%%%%%%%%%%%%%%%%%%%%%%%%%%%%%%%%%%%%%%%%%%%%%%%%%%%%%%%%%%%%%%%%%%%%%

\begin{abstract}

Broad infrared emission features (e.g., at 3.3, 6.2, 7.7, 8.6, and 11.3~\microns) from the 
gas phase interstellar medium have long been attributed to polycyclic aromatic 
hydrocarbons (PAHs). A significant portion (10\%$-$20\%) of the Milky Way's carbon 
reservoir is locked in PAH molecules, which makes their characterization integral to our 
understanding of astrochemistry. In molecular clouds and the dense envelopes and disks of young 
stellar objects (YSOs), PAHs are expected to be frozen in the icy mantles of dust grains where they 
should reveal themselves through infrared absorption. To facilitate the search for frozen interstellar 
PAHs, laboratory experiments were conducted to determine the positions and strengths of the 
bands of pyrene mixed with \h2o\ and \d2o\ ices. The \d2o\ mixtures are used to measure 
pyrene bands that are masked by the strong bands of \h2o, leading to the first laboratory 
determination of the band strength for the CH stretching mode of pyrene in water ice near 
3.25~\microns. Our infrared band strengths were normalized to experimentally determined 
ultraviolet band strengths, and we find that they are generally $\sim$50\% larger than 
those reported by \citeauthor{2011A&A...525A..93B}\ based on theoretical strengths. These improved band 
strengths were used to reexamine YSO spectra published by \citeauthor{2008ApJ...678..985B}\ 
to estimate the contribution of frozen PAHs to absorption in the 5$-$8~\microns\ spectral region, 
taking into account the strength of the 3.25~\microns\ CH stretching mode. It is found that frozen 
neutral PAHs contain 5\%$-$9\% of the cosmic carbon budget, and account for 2\%$-$9\% of the unidentified 
absorption in 
the 5$-$8~\microns\ region.

\end{abstract}

%%%%%%%%%%%%%%%%%%%%%%%%%%%%%%%%%%%%%%%%%%%%%%%%%%%%%%%%%%%%%%%%%%%%%%%%%%%%

\section{Introduction}

Unidentified infrared bands discovered almost 40 yr ago in the interstellar medium (ISM) are now
attributed to the CH and CC vibrational modes of polycyclic aromatic hydrocarbons (PAHs). They are 
estimated to account for 10\%$-$20\% of the total carbon reservoir in the ISM \citep{2011IAUS..280..149P}.
%\citep{1992ApJ...393L..79J,1999A&A...352..659A,2002nla..work..131H} 
Absorption features at 3.25 \citep{1994ApJ...433..179S,1996ApJ...459..209B,1999ApJ...517..883B}, 6.2 \citep{2001A&A...376..254K}, and 11.3 \microns\ \citep{2000ApJ...544L..75B} have been attributed 
to PAHs in a handful of young stellar object (YSO) spectra, while PAH emission bands are weak or absent toward embedded YSOs 
\citep{2009A&A...495..837G}. In molecular clouds and in the disks and envelopes of embedded YSOs, PAHs, along with 
other gas phase species, are expected to freeze out on the icy mantles of dust grains. 
In the dense ISM, ice mantles
are important sites for prebiotic chemistry, where the long residence times of condensed molecules give 
them the best opportunity for interaction. Indeed, laboratory experiments have shown the ease of 
producing biomolecules via irradiation and heating of interstellar ice analogs \citep[e.g.][]{2002Natur.416..401B, 2002Natur.416..403M,2013ApJ...765..111K}. Much work has been done to understand PAH chemistry in ice 
matrices \citep[e.g.][]{2003ApJ...596L.195G,2004JPC...108.4412G,2004ApJ...615L.177G}. In particular, PAHs 
frozen in amorphous \h2o\ ice ionize quickly when irradiated by UV light. Larger PAHs remain stable 
up to 120~K, at which point the ice crystallizes and PAHs begin to react with the ice matrix itself \citep{2006ApJ...638..286G,2011EAS....46..305A}, while smaller PAH ions react at lower temperatures 
to form hydrogenated and oxygenated complex molecules \citep{2012ApJ...756L..24G}. 

Significant attention has also been given toward the task of quantifying PAHs in the ISM. 
An extensive online database at www.astrochem.org \citep{2010ApJS..189..341B,2014BoersmaArticle} contains both theoretical and experimental spectra of PAHs matrix-isolated in argon. These 
spectra greatly enhance our understanding of PAH signatures, but spectra of PAHs embedded in interstellar 
ice analogs are necessary to truly constrain PAH abundances within the dense ISM. Experiments 
have been carried out to determine band strengths for both neutral 
\citep{2005ApJS..161...53B,2011A&A...525A..93B} and ionized \citep{2007ApJ...664.1264B} 
pyrene frozen in water ice, but the absolute infrared (IR) band strengths reported in these cases were 
normalized to theoretically calculated band strengths. 
Sometimes, the region of interest reported was constrained between 
1650 to 1000~cm$^{-1}$ (6$-$10~\microns) because the \h2o\ ice absorption 
there is less significant and more linear than in other regions, and none of these studies address the CH 
stretching mode at 3.25~\microns, which is most easily accessible with ground-based telescopes. 
Band strengths for naphthalene frozen in water ice have been reported as well \citep{2004ApJ...607..346S}.

The aim of our work is to improve upon previous band strength estimates by recording PAH spectra in 
both infrared and ultraviolet-visible (UV-Vis) bands, then normalizing the results to previous, direct 
measurements of UV absolute band strengths \citep{Berlman1971, Dixon2005}. This paper reports our 
results for pyrene (C$_{16}$H$_{10}$) frozen in either \h2o\ or \d2o\ ice. Pyrene was chosen as a 
representative PAH molecule because 
it has been widely studied in the laboratory and a significant amount of laboratory data exists.
The pyrene radical cation is proposed to be the carrier of some of the diffuse interstellar bands (DIBs) 
\citep{1992Natur.358...42S}. Additionally, pyrene is the smallest available PAH that is convenient to 
handle but at the same time compact with a C/H ratio of 1.9 which is closer to the larger PAH C/H ratio 
expected in the ISM (see Section~\ref{sec:carbon}).
We included \d2o\ in our experiments because its absorption features are redshifted compared 
to those in \h2o, allowing easier measurement and detection of some pyrene bands which would 
normally be drowned out by \h2o\ features. With 
revised band strengths in hand, we attempted to identify the previously reported 
3.25~\microns\ band \citep{1994ApJ...433..179S,1996ApJ...459..209B,1999ApJ...517..883B} 
in the set of YSO spectra 
published by \citet{2008ApJ...678..985B} and, where found, constrained PAH column densities and the 
contribution of PAH absorption to the 5$-$8~\microns\ absorption region. Section 
\ref{sec:experiment} describes the experimental setup for obtaining spectral measurements while Section 
\ref{sec:analysis} details our procedure for calculating absolute band strengths. Section 
\ref{sec:ysos} describes the astrophysical implications for quantifying PAHs in the dense ISM. Section 
\ref{sec:summary} summarizes our findings and 
suggests possible trajectories for further expansion of this work.

%%%%%%%%%%%%%%%%%%%%%%%%%%%%%%%%%%%%%%%%%%%%%%%%%%%%%%%%%%%%%%%%%%%%%%%%%%%%

\section{Experimental Setup}
\label{sec:experiment}

\begin{figure}[H] 
	\centering 
	\includegraphics[angle=0,width=.8\textwidth]{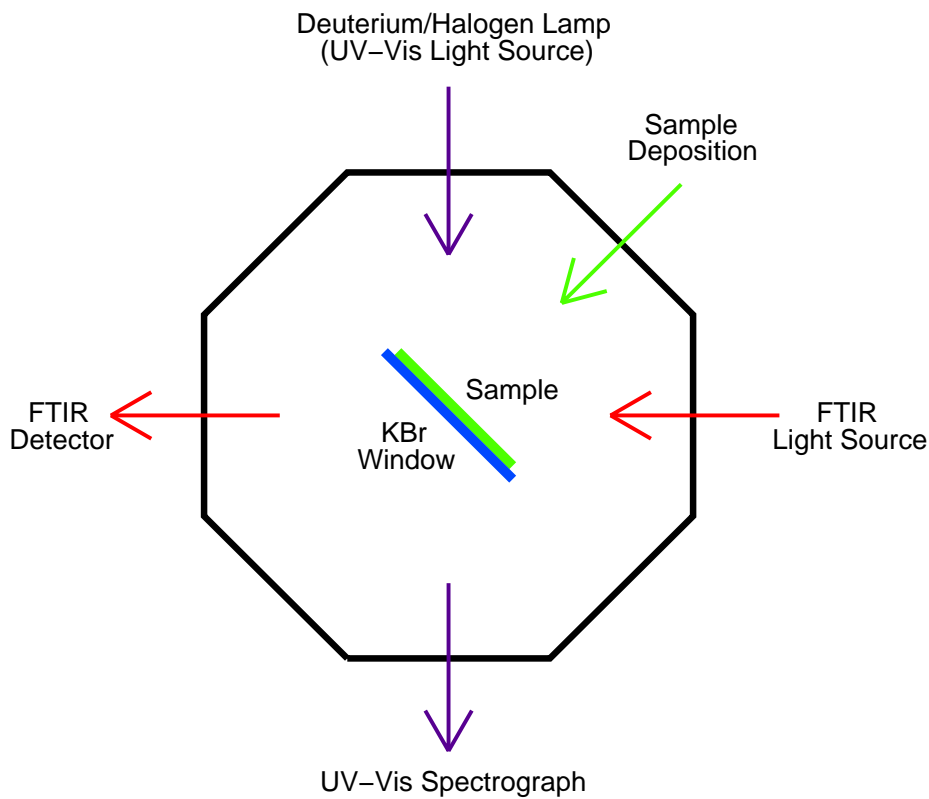} 
	\caption{Experimental setup. The pressure in the chamber was kept at $\sim$10$^{-7}$~mbar, 
	and the temperature of the KBr window was kept at 30 K. The frozen pyrene sample was generated 
	by passing water vapor over sublimating pyrene crystals. 
	This gaseous mixture then froze onto the KBr window. Transmission spectra were recorded using 
	FTIR and UV-Vis spectrometers.} 
	\label{fig:cartoon} 
\end{figure}

The setup for our laboratory experiment was very similar to that described in 
\citet{2012ApJ...747...13B}; a schematic is shown in Figure \ref{fig:cartoon}. In summary, the chamber for this 
experiment was cooled to 30~K at which temperature the base pressure in the chamber was in the 
high $10^{-9}$ to low $10^{-8}$~mbar. Water vapor was passed over pyrene 
crystals which were heated to $\sim$55$^{\circ}$C for sublimation under 
vacuum. This mixture of sublimated pyrene and water vapor was directed into the chamber, where 
it subsequently froze onto a potassium bromide (KBr) window. During the ice 
deposition, pressure in the chamber was typically around 10$^{-7}$ to 10$^{-6}$ mbar due to the 
water vapor lead into the chamber. Under these 
conditions, the pyrene concentration was less than three percent in all samples, ensuring that 
pyrene molecules were adequately isolated from each other within the ice matrix. The pyrene concentration 
was further monitored and controlled as necessary during the experiments by measuring the ice film 
thickness, adjusting the pyrene sublimation temperature, and monitoring the IR and 
UV-Vis absorption features. Ice film deposition rates were approximately 1~\microns\ per hour. Transmission spectra through samples ranging in thickness from a few hundred nanometers to 5~\microns\ were 
collected via Fourier transform infrared (FTIR) spectroscopy and a UV-Vis spectrograph using a 
deuterium/halogen lamp as a continuum source. The spectral resolution and spectral coverage of 
the FTIR spectrometer were 1~cm$^{-1}$ and 8000$-$400~cm$^{-1}$, respectively. The UV-Vis spectrometer 
had a resolution of $\sim$0.75~nm in the 210$-$1100~nm wavelength range. Each instrument was 
connected to a separate port of the octagonal chamber, allowing IR and UV spectra to be collected 
consecutively. The optical window was oriented in such a way that it was as close to perpendicular 
as possible to the deposition side, at the same time allowing both FTIR and UV-Vis spectral measurements 
without rotating the cryostat on which the optical window was mounted. This configuration gave us highly 
reproducible spectra as we grew the ice slowly on the optical window. Roughly half the experiments were 
conducted with \h2o\ and half with \d2o\ in order to recover better measurements in areas with 
significant ice absorption. During \d2o\ matrix experiments, even though we passivated the 
chamber with \d2o, there was still, at times, obvious contamination of \h2o\ ice features 
in the spectra, necessitating great care when interpreting band strengths in regions of overlap between ice features.

Atmospheric water vapor in the path of the FTIR instrument and the detector was minimized by 
passing gaseous nitrogen through the enclosed regions. However, there were several openings, 
which were necessary to optimize the instrumentation, that could not be completely sealed off. As 
a consequence, vapor-phase water absorption is always present in our infrared spectra. 
Fortunately, due to their characteristic absorption, we have routinely subtracted the water vapor 
absorption from the ice spectra, as will be discussed in the next section. Water vapor spectra were 
derived from measuring the spectra of the experimental set-up at room temperature before and 
after passing dry nitrogen gas at a very high flow rate (10 times more than the normal flow rate of 
$\sim$20~L/minute), eliminating a significant amount of water vapor over a short period of time.

%%%%%%%%%%%%%%%%%%%%%%%%%%%%%%%%%%%%%%%%%%%%%%%%%%%%%%%%%%%%%%%%%%%%%%%%%%%%

\section{Analysis}
\label{sec:analysis}

Absolute IR band strengths of ices are difficult to calculate directly from laboratory experiments, because 
column densities require a determination of the sample density, thickness, and mixing ratio. In order to 
determine absolute IR band strengths of pyrene frozen in water ice, we correlated our measured IR pyrene 
bands to the first electronic transition of pyrene. The absolute band strength of this 
UV band is easily calculated from previously published data.

\subsection{UV Band Strength And Integrated Absorption}

The molar absorptivity, $\epsilon$ (M$^{-1}$~cm$^{-1}$), of the first electronic transition of 
pyrene (in cyclohexane) was measured by \citet{Berlman1971}. This transition occurs in the UV 
frequency range of $\sim$35,000$-$28,800~cm$^{-1}$ (286$-$347~nm). \citet{Dixon2005} 
duplicated the experiment and published their results in the online database 
PhotochemCAD.\footnote{www.photochemcad.com} Their results were normalized at the peak 
of the first electronic transition to the value $\epsilon=54,000$~M$^{-1}$~cm$^{-1}$ listed by 
\citet{Berlman1971}, which agrees with the results found by \citet{1997PCPB..104..25T} for gas 
phase pyrene at 150~K. From these measurements, we directly calculated the integrated band strength of the first electronic transition of pyrene using Equations \eqref{eq:band}-\eqref{eq:bandep}. 
Although these experiments measured pyrene in cyclohexane rather than in ice, the total oscillator strength of any given electronic transition remains constant (a molecular property), irrespective of the solvents used, unless strong electronic interactions between the solvent and the PAH molecule occurs that could change electronic and vibrational transitions significantly. This is not the case in cyclohexane which is a non-polar solvent. Amorphous ices behave like non-polar solvents with electronic polarizability similar to argon matrix, so strong interactions between pyrene and amorphous water ice are negligible \citep{2004JPC...108.4412G}. Therefore, the assumption that the integrated UV band strength for the first electronic transition of pyrene is equivalent in amorphous water ice and cyclohexane is valid.

Integrated band strengths, $A$, are calculated using 
\begin{equation}
A=\frac{\int_{x_{1}}^{x_{2}} \tau(x) dx}{N}=\frac{2.303\int_{x_{1}}^{x_{2}}
a(x) dx}{N}
\label{eq:band}
\end{equation}
where $\tau=ln(I_0/I)$ is the
optical depth of the feature, $x$ is the wavenumber ($1/\lambda$, cm$^{-1}$), 
$N$ is the column density (cm$^{-2}$) of the material producing the feature, and 
$a=log_{10}(I_0/I)$ stands for absorbance, which is the preferred unit of experimental 
scientists.

After some manipulation, the Beer$-$Lambert law \eqref{eq:beer} can be substituted into
\eqref{eq:band} to determine the band strength directly from molar absorptivity.
In Equations \eqref{eq:beer}-\eqref{eq:bandep}, $c$ is molar concentration 
(1 M = 0.001 mol~cm$^{-3}$), $l$ is path length (cm), $n$ is number density 
(cm$^{-3}$), and $N_A$ is Avogadro's constant ($6.02\times10^{23}~\rm{mol}^{-1}$).

\begin{equation}
	a=\epsilon c l
\label{eq:beer}
\end{equation}

\begin{equation}
	N=nl=\left(0.001 \left[\frac{\rm{mol~cm}^{-3}}{M}\right]\right) N_A c l
\label{eq:n}
\end{equation}

\begin{equation}
	A=\frac{2.303\int_{x_{1}}^{x_{2}} \epsilon(x) dx}{\left(0.001\left[\frac{\rm{mol~cm}^{-3}}{M}\right]\right)N_A}
\label{eq:bandep}
\end{equation}

\begin{figure}[H]
	\centering
	\includegraphics[angle=0,scale=.75]{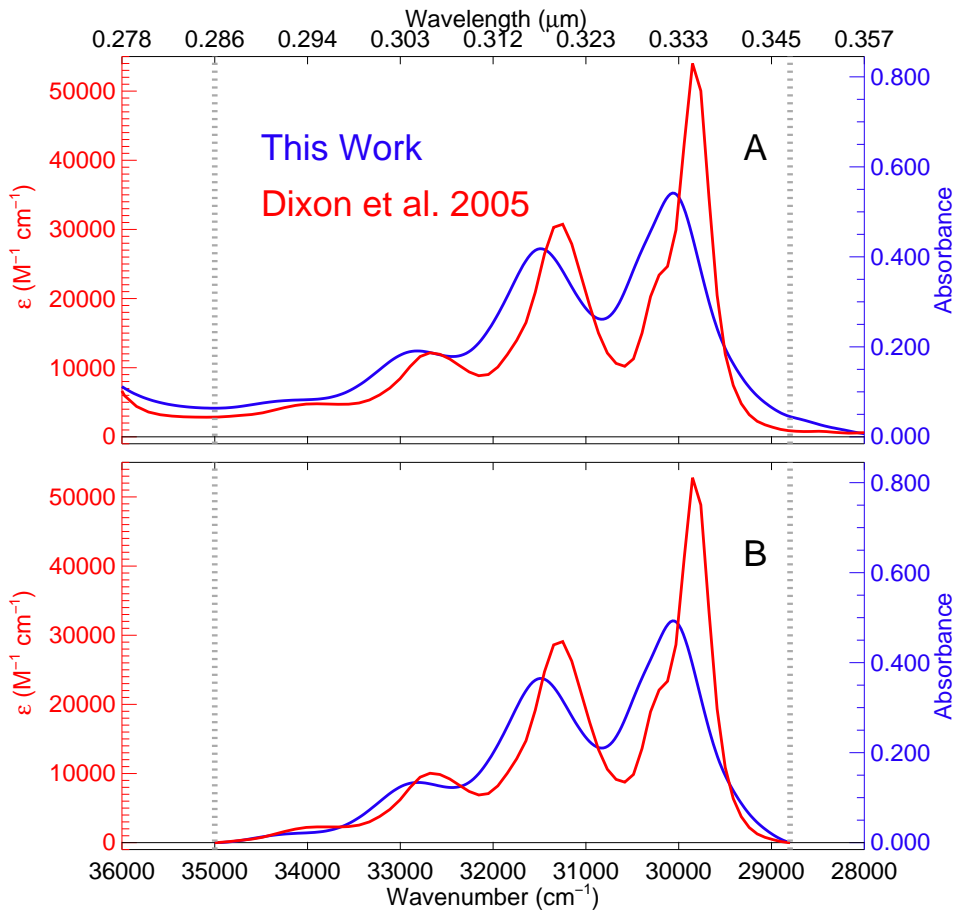}
	\caption{Panels A and B show intermediate steps for obtaining the absolute UV band strength for 
the first electronic transition of pyrene. An absorption spectrum selected from our sample of pyrene in \h2o\ is shown 
in blue. The molar absorptivity spectrum of pyrene in cyclohexane, published by \citet{Dixon2005}, is 
shown in red. Panel A shows original spectra, without baselining. Panel B shows the same spectra, after 
subtracting linear baselines anchored at 35,000 and 28,800 cm$^{-1}$ (dotted lines).}
	\label{fig:UVabsband}
\end{figure}

Before performing any of the following analysis, raw spectra were smoothed using a fast Fourier transform (FFT) filter \citep{BowmanIDL} with a window\footnote{See http://wiki.originlab.com/$\sim$originla/wiki2/index.php?title=X-Function:Smooth. Note, however, a typo in the equation listed there. The term $(1-t)^2$ in the last line of the equation should instead be $(n-t)^2$.} of $p=5$ in order not to compromise actual signal.
% described in Equation \eqref{eq:fft} to remove some of the experimental noise. 
% Add in thesis
%http://www.originlab.com/www/helponline/origin/en/Programming/mergedProjects/X-Function/X-Function/Smooth.html
%\begin{equation}
%		w(i)=
%			\begin{cases}
%			1-\frac{4p^2}{n^2}i^2, & 0 \leq i \leq \frac{n}{2p}\\
%			0, & \frac{n}{2p} < i < n-\frac{n}{2p}\\
%			1-\frac{4p^2}{n^2}(n-i)^2, & n-\frac{n}{2p} \leq i \leq n\\
%			\end{cases}
%\label{eq:fft}
%\end{equation}
% Take spectrum in spatial domain, FFT to frequency domain, remove intermediate frequencies
% via the filter, transform back to spatial domain
The integration of molar absorptivity was carried out after subtracting a local, linear baseline
anchored at 35,000 and 28,800~cm$^{-1}$ (see Figure \ref{fig:UVabsband}). We calculated the UV band strength as 
$2.578\times10^{-13}$~cm/molecule. %Although the behavior of pyrene in cyclohexane is slightly different from its behavior in water ice due to matrix interactions (see Figure \ref{fig:UVabsband}), the integrated band strength of the feature should remain the same. Since the peak of the pyrene transition in cyclohexane is shifted by $\sim$~3~nm from the peak in ice, we shifted the integration limits stated above $\pm$~3~nm to estimate the error in the band strength calculation due to the choice of baseline, which we found to be only a few percent.

The errors in molar extinction coefficients come not from the spectra, but from determining the weight of 
the substance in solution (molarity) as well as whether the molecules form aggregates in solution or 
stay as monomers \citep{SiuandDuhamel2008}. A good comparative study on absorption of pyrene 
in various solvents is given in \citet{Rayetal2006}, where significant changes in peak positions
are seen. However, their integrated molar absorptivity remains essentially unaltered, because if the 
height decreases, the bandwidth increases. Hence, we estimate that the experimental band strengths 
derived for pyrene UV spectra are very accurate within a few percent uncertainty at the maximum.

Integrated absorbances of the first electronic transition of pyrene were calculated from our spectra
using the same linear baseline as described above. Subtracting a local baseline is important for frozen
pyrene, because absorption due to the ice matrix itself is significant. Errors on the integrated UV 
absorbances were determined by shifting the baseline by 3~nm in either direction. In the rest of this work, the pyrene column densities calculated from the UV spectra were used to derive band strengths for the infrared transitions.

\subsection{IR Integrated Absorption}

\begin{figure}[H]
	\centering
	\includegraphics[angle=0,scale=.8]{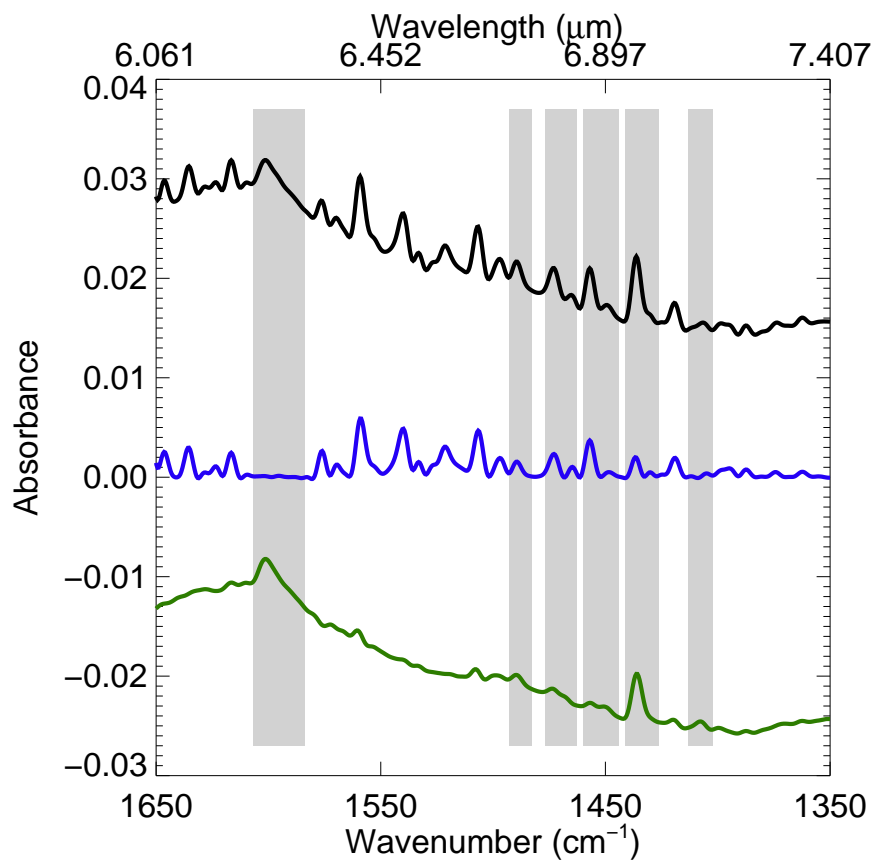}
	\caption{\small Typical example of water vapor subtraction. Regions where pyrene features exist are marked in gray. The original spectrum is shown in black, and the corrected spectrum (green) after subtraction of water vapor contamination (blue) is offset below.}
\label{fig:wv}
\end{figure}

To determine the strength, peak position, and width of the IR pyrene features, contamination by \h2o\ vapor 
and baseline curvatures (due to ice absorption) need to be removed first. Contamination by water vapor 
absorption is significant in many of the spectra from 
$\sim$2025$-$1275~cm$^{-1}$ (see Figure~\ref{fig:wv}). Without removal, such contamination greatly affects the 
identification and analysis of almost half of the IR pyrene bands. To obtain a water vapor template, we took 
the ratio of two background spectra. The water vapor residual was cleaned by the same FFT smoothing 
described above, and a local, low-order polynomial baseline was subtracted. %(third/fifth)-order
The amplitudes of the five largest water vapor peaks unaffected by ice absorption were measured in each 
spectrum, and the water vapor template, scaled to those amplitudes, was then subtracted to remove the 
contamination. 

\begin{figure}[H]
	\centering
	\includegraphics[angle=0,scale=.31]{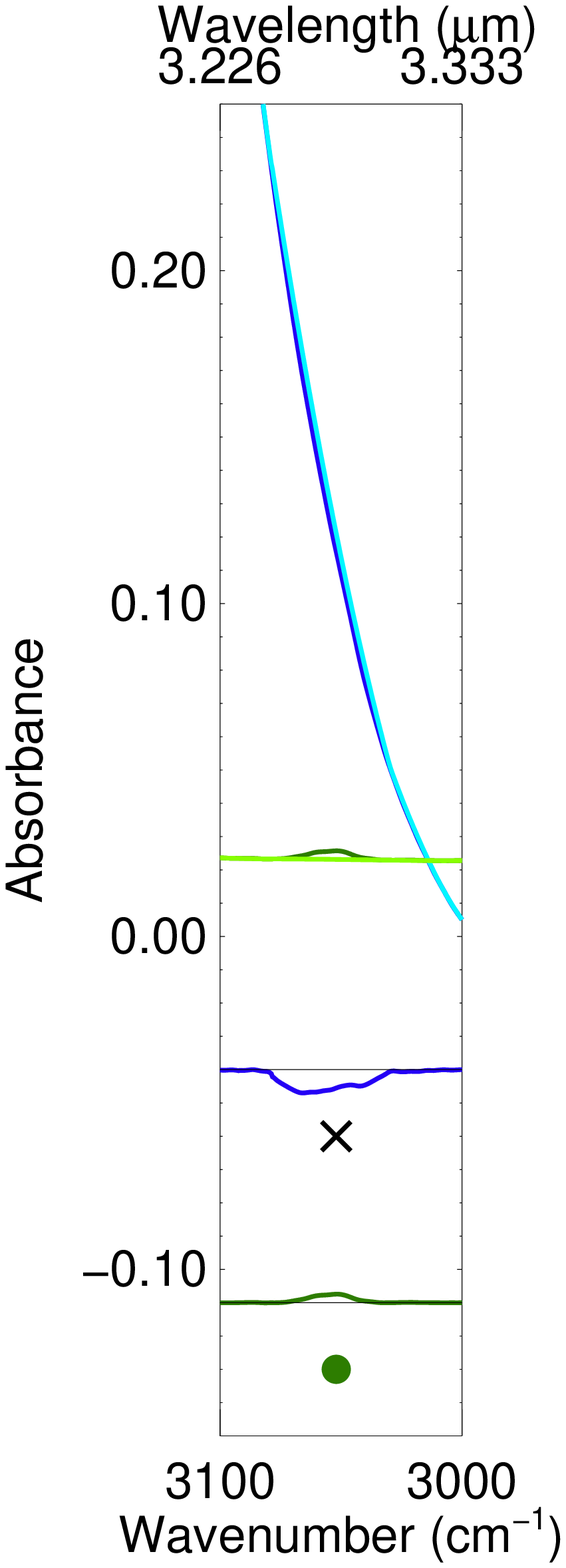}
	\includegraphics[angle=0,scale=.31]{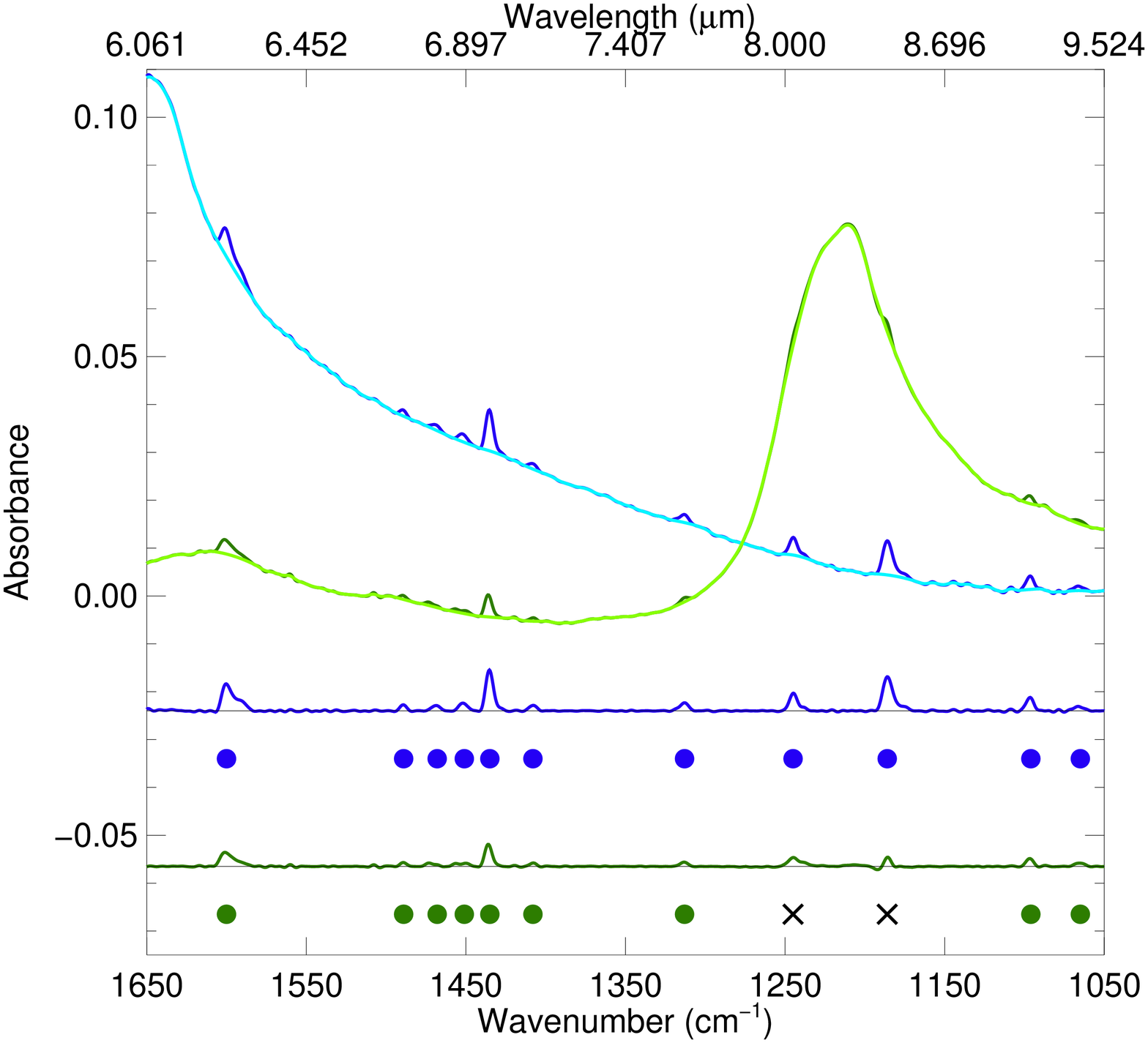}
	\includegraphics[angle=0,scale=.31]{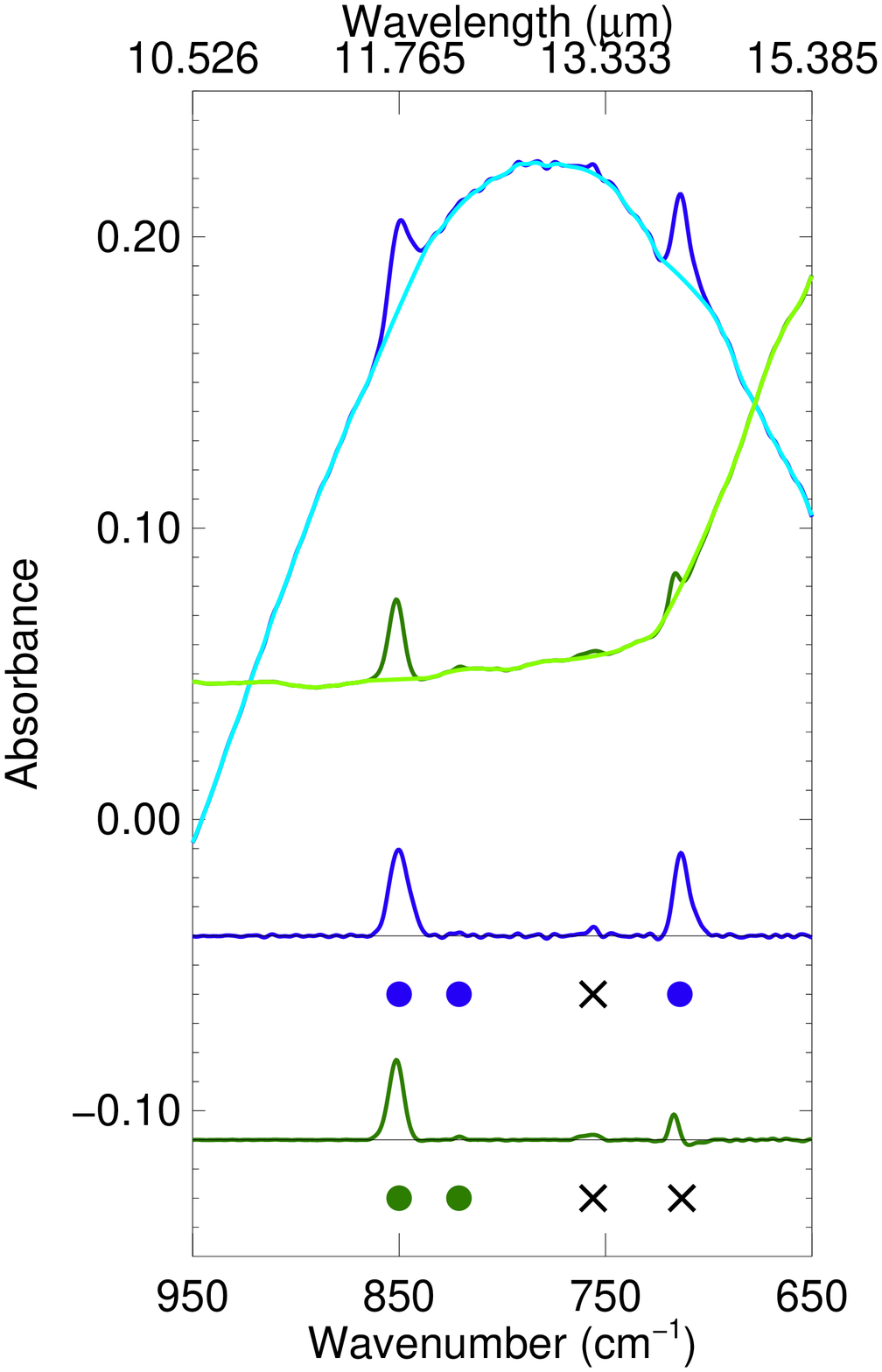}
	\caption{\small Typical examples of pyrene/\h2o\ and pyrene/\d2o\ spectra, after water vapor removal, are plotted with their baselines in the regions of pyrene absorption. \h2o is in blue, and \d2o is in green. The PAH residual spectra are offset below. Due to the way in which features were masked during baselining, not all PAH features were measured in each spectrum. Features that were measured are marked with a filled circle, and features that were ignored are marked with a cross. Note that the central panel has a much smaller $y$-axis scaling.}
\label{fig:baseline}
\end{figure}

To determine the integrated absorption of pyrene features in the IR, a global baseline to remove ice
absorption was determined through implementation of the following method. First, pyrene 
features were masked (see Table \ref{tab:mask}), and the remaining spectrum was smoothed using 
Gaussian convolution. 
Then, the smoothed spectrum was interpolated across feature regions using a second-order 
polynomial. This global baseline was subtracted from the original spectrum to obtain a residual 
pyrene spectrum for which position, FWHM, and integrated area was determined for each band. Figure 
\ref{fig:baseline} provides an example of this method.

\begin{deluxetable}{cc}
\tablecolumns{2}
\tablewidth{0pt}
\tablecaption{Masking Regions for Global Continuum
	\label{tab:mask}}
\tablehead{
\colhead{Band Identification} & \colhead{Masking Region (cm$^{-1}$)}}
\startdata
	A          & 3078$-$3032       \\
	B          & 1607$-$1584       \\
	C          & 1493$-$1483       \\
	D          & 1477$-$1463       \\
	E          & 1460$-$1444        \\
	F          & 1441$-$1426        \\
	G          & 1413$-$1402       \\
	H          & 1321$-$1308       \\
	I           & 1251$-$1235        \\
	J           & 1194$-$1171        \\
	K          & 1103$-$1091        \\
	L           & 1072$-$1059       \\
	M          & 864$-$837           \\
	N          & 826$-$817           \\
	O          & 767$-$746           \\
	P           & 722$-$700           \\
\enddata
\tablecomments{For this analysis, features were manually selected by visual inspection of the laboratory spectra using \citet{2011A&A...525A..93B} and the experimental results for pyrene in argon from www.astrochem.org \citep{2010ApJS..189..341B,2014BoersmaArticle} as a guide. Although an additional feature at 1136~\wavenum\ was 
reported by \citet{2011A&A...525A..93B}, it was not detected above 3$\sigma$ in our sample.}
\end{deluxetable}

The error in the integrated area of IR bands was calculated by taking into account the standard
deviation of absorption in the pyrene residuals. Since the noise can vary slightly with wavenumber, 
the standard deviation in absorption for a given feature was calculated within a region $\pm$3 times 
the width of the masking region for that feature, ignoring any other pyrene features within 
these limits.

%\clearpage
%\begin{landscape}
%\end{landscape}
%\clearpage

\subsection{Absolute IR Band Strengths}

\begin{figure}[H]
	\centering
	\includegraphics[angle=0,scale=.6]{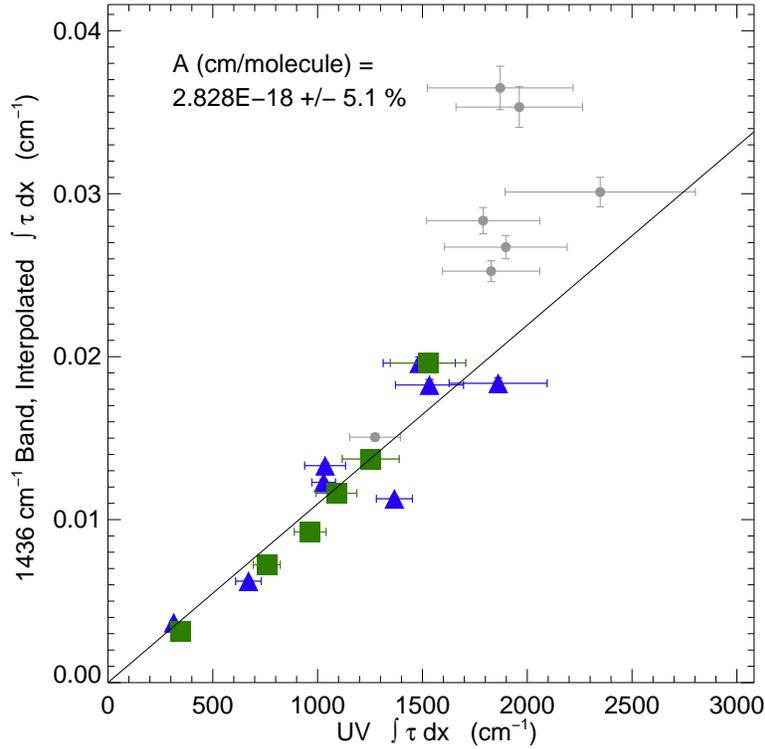}
	\caption{Correlation between the 1436~cm$^{-1}$ IR band and the 35,000$-$28,800~cm$^{-1}$ UV band. 
Triangles show \h2o\ samples, and squares show \d2o\ samples. Small circles show 
how the correlation breaks down as the UV band begins to saturate. Non$-$linear growth of the UV features 
becomes significant above a peak absorbance of 0.8 (or $\tau$ of 1.84). Only unsaturated data points were used to 
determine the best fit linear correlation with the condition that it passes through the origin.}
	\label{fig:UVvsF}
\end{figure}

The absolute IR band strengths were determined by correlating a strong infrared band with the 
calibrated UV band. This was challenging because the UV bands of pyrene tend to saturate fairly 
quickly while the weaker IR bands can take much longer to build up enough signal-to-noise for 
measurement. Figure \ref{fig:UVvsF} shows the correlation of the 1436 cm$^{-1}$ band, as shown in Figure \ref{fig:baseline}, with the UV band. This band is both fairly strong and relatively unaffected by ice absorption in either 
\h2o\ or \d2o ice. Since IR and UV spectra were taken consecutively rather than concurrently, IR measurements 
had to be interpolated to the observation times of the UV spectra. After determining the absolute band strength 
for the 1436 cm$^{-1}$ band, the strengths of all bands were determined by correlating them with each other. 
Linear correlations were determined using the IDL program MPFITEXY \citep{2010MNRAS.409.1330W} which 
utilizes the MPFIT package \citep{2009ASPC..411..251M}. MPFIT 
is a curve fitting program that uses a robust, non$-$linear least squares method. MPFITEXY was used to 
take into account $x$ and $y$ error bars and force the $y$-intercepts through the origin. Table \ref{tab:final} lists 
the position and FWHM of each band averaged over all spectra along with the final calculated band strengths. 

\begin{deluxetable}{ccccccccc}
\tabletypesize{\scriptsize}
\rotate
\tablecolumns{9}
\tablewidth{0pt}
\tablecaption{IR Pyrene Band Characteristics
	\label{tab:final}}
\tablehead{
\multicolumn{6}{c}{This Work} & \colhead{} & \multicolumn{2}{c}{\citet{2011A&A...525A..93B}}\\
\cline{1-6} \cline{8-9} \\
\colhead{ID} & \colhead{Position} & \colhead{FWHM} & \colhead{Position} & \colhead{FWHM} & \colhead{$A$} &\colhead{} & \colhead{Position} & \colhead{$A^{\rm{\textbf{a}}}$} \\
\colhead{} & \colhead{\wavenum} & \colhead{\wavenum} & \colhead{\microns} & \colhead{\microns} & \colhead{10$^{-19}$ cm/molecule} &\colhead{} & \colhead{\wavenum} & \colhead{10$^{-19}$ cm/molecule}}
\startdata
A$^{\rm{\textbf{b}}}$ & 3051.98 $\pm$ 0.41 & 20.50 $\pm$ 0.17 & 3.2766 $\pm$ 0.0004 & 0.0220 $\pm$ 0.0002 & 60.36 $\pm$ 3.12 & & $\cdots$ & $\cdots$ \\
B & 1600.47 $\pm$ 0.05 & 8.83 $\pm$ 0.07 & 6.2482 $\pm$ 0.0002 & 0.0345 $\pm$ 0.0003 & 33.56 $\pm$ 1.73 & & 1600.5, 1594.1$^{\rm{\textbf{c}}}$ & 12.0, 12.7\\
C & 1488.79 $\pm$ 0.06 & 5.28 $\pm$ 0.06 & 6.7169 $\pm$ 0.0003 & 0.0238 $\pm$ 0.0003 & 3.10 $\pm$ 0.17 & & 1488.4 & 3.3 \\
D & 1468.42 $\pm$ 0.11 & 6.76 $\pm$ 0.24 & 6.8101 $\pm$ 0.0005 & 0.0313 $\pm$ 0.0011 & 3.80 $\pm$ 0.21 & & 1468.4 & 1.8\\
E & 1451.52 $\pm$ 0.15 & 8.02 $\pm$ 0.21 & 6.8893 $\pm$ 0.0007 & 0.0380 $\pm$ 0.0010 & 6.59 $\pm$ 0.35 & & 1452.0 & 2.8 \\
F & 1435.60 $\pm$ 0.04 & 5.71 $\pm$ 0.06 & 6.9658 $\pm$ 0.0002 & 0.0277 $\pm$ 0.0003 & 28.28 $\pm$ 1.44 & & 1435.1 & 17.3 \\
G & 1407.86 $\pm$ 0.04 & 5.23 $\pm$ 0.05 & 7.1030 $\pm$ 0.0002 & 0.0264 $\pm$ 0.0003 & 3.71 $\pm$ 0.20 & & $\cdots$ & $\cdots$ \\
H & 1313.49 $\pm$ 0.06 & 6.12 $\pm$ 0.08 & 7.6133 $\pm$ 0.0003 & 0.0355 $\pm$ 0.0005 & 6.41 $\pm$ 0.34 & & 1313.7 & 5.3 \\
I$^{\rm{\textbf{a}}}$ & 1244.84 $\pm$ 0.18 & 6.48 $\pm$ 0.16 & 8.0332 $\pm$ 0.0011 & 0.0418 $\pm$ 0.0010 & 13.08 $\pm$ 0.68 & & 1244.0 & 10.3 \\
J$^{\rm{\textbf{a}}}$ & 1186.10 $\pm$ 0.11 & 7.31 $\pm$ 0.20 & 8.4310 $\pm$ 0.0008 & 0.0520 $\pm$ 0.0014 & 34.24 $\pm$ 1.76 & & 1185.6,1176.3$^{\rm{\textbf{c}}}$ & 21.7, 6.1 \\
K & 1096.46 $\pm$ 0.05 & 5.33 $\pm$ 0.07 & 9.1203 $\pm$ 0.0004 & 0.0443 $\pm$ 0.0006 & 7.62 $\pm$ 0.41 & & 1096.4 & 5.1 \\
L & 1066.07 $\pm$ 0.19 & 7.20 $\pm$ 0.27 & 9.3802 $\pm$ 0.0017 & 0.0634 $\pm$ 0.0024 & 3.24 $\pm$ 0.25 & & 1065.5 & 2.8 \\
M$^{\rm{\textbf{a}}}$ & 850.51 $\pm$ 0.25 & 11.62 $\pm$ 0.21 & 11.7578 $\pm$ 0.0034 & 0.1609 $\pm$ 0.0029 & 191.25 $\pm$ 9.96 & & $\cdots$ & $\cdots$ \\
M$^{\rm{\textbf{b}}}$ & 850.90 $\pm$ 0.11 & 9.31 $\pm$ 0.18 & 11.7523 $\pm$ 0.0016 & 0.1288 $\pm$ 0.0025 & 313.51 $\pm$ 16.18 & & $\cdots$ & $\cdots$ \\
N & 821.11 $\pm$ 0.07 & 4.88 $\pm$ 0.09 & 12.1787 $\pm$ 0.0011 & 0.0724 $\pm$ 0.0013 & 7.05 $\pm$ 0.42 & & $\cdots$ & $\cdots$ \\
P$^{\rm{\textbf{a}}}$ & 714.01 $\pm$ 0.09 & 8.65 $\pm$ 0.16 & 14.0055 $\pm$ 0.0018 & 0.1700 $\pm$ 0.0031 & 140.73 $\pm$ 7.30 & & $\cdots$ & $\cdots$ \\
\enddata
\vspace{-0.8cm}
\tablecomments{Some bands could only be measured in one ice matrix due to the inability to accurately determine the baseline ice absorption in the other ice species. Band M was measured separately in both \h2o\ and \d2o\ ice, as it appears to be $\sim$50\% larger in a \d2o\ ice matrix. $^{\rm{\textbf{a}}}$Measured only in \h2o. $^{\rm{\textbf{b}}}$Measured only in \d2o. $^{\rm{\textbf{c}}}$Our laboratory spectra show one feature at this wavelength, whereas \citet{2011A&A...525A..93B} report two.}
\end{deluxetable}

\subsubsection{Comparison of Pyrene Band Strengths in \h2o\ and \d2o\ Ices}

\begin{figure}[H]
	\centering
	\includegraphics[angle=90,width=\textwidth]{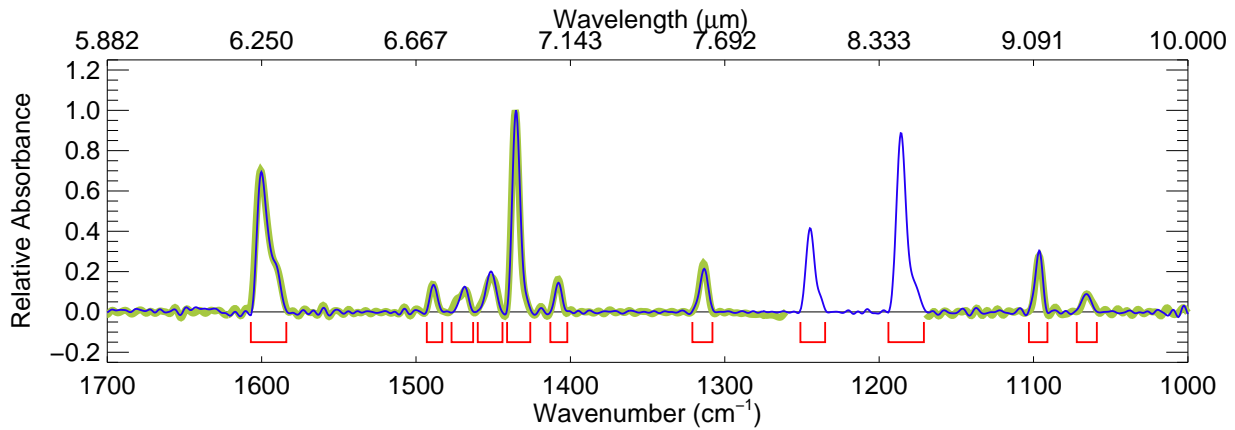}
	\caption{Comparison of the pyrene residual spectrum averaged across \h2o\ (blue) and \d2o\ 
(green) experiments. The average spectra were normalized at 1436~\wavenum. Regions of obvious 
pyrene features are marked below in red. Some regions had to be excluded from the analysis due to 
inability to accurately determine baseline ice absorption, e.g., near 1200~\wavenum\ in \d2o\ ice.}
	\label{fig:h2od2o}
\end{figure}

Spectra were recorded for pyrene in \d2o\ as well as in \h2o ice in order to recover bands masked 
in laboratory spectra by the strong \h2o\ features (e.g., the 3.25~\microns\ CH stretching mode). 
This requires an assumption that CH stretching modes of pyrene 
behave identically whether the pyrene is frozen in \h2o\ or \d2o\ ice. Figure~\ref{fig:h2od2o} 
demonstrates that this assumption is valid at least for the CC stretching and CH in plane bending modes of pyrene. It is not clear that this is the case for the CH out of plane bending modes. Unfortunately, beyond 10~\microns, ice absorption in both \h2o\ and \d2o is quite strong and makes characterization of pyrene bands in that region more difficult. Local baselines could not be adequately determined for many of the pyrene bands beyond 10~\microns, but the 850~\wavenum\ band appears to be measurable in both ice matrices. This was the only band in our analysis which was measured in both ices and gave different results depending on the ice matrix, as noted in Table~\ref{tab:final}.

\subsection{Comparison With Previous Laboratory Work}

\begin{figure}[H]
	\centering
	\begin{tabular}{cc}
		\includegraphics[angle=0,scale=.45]{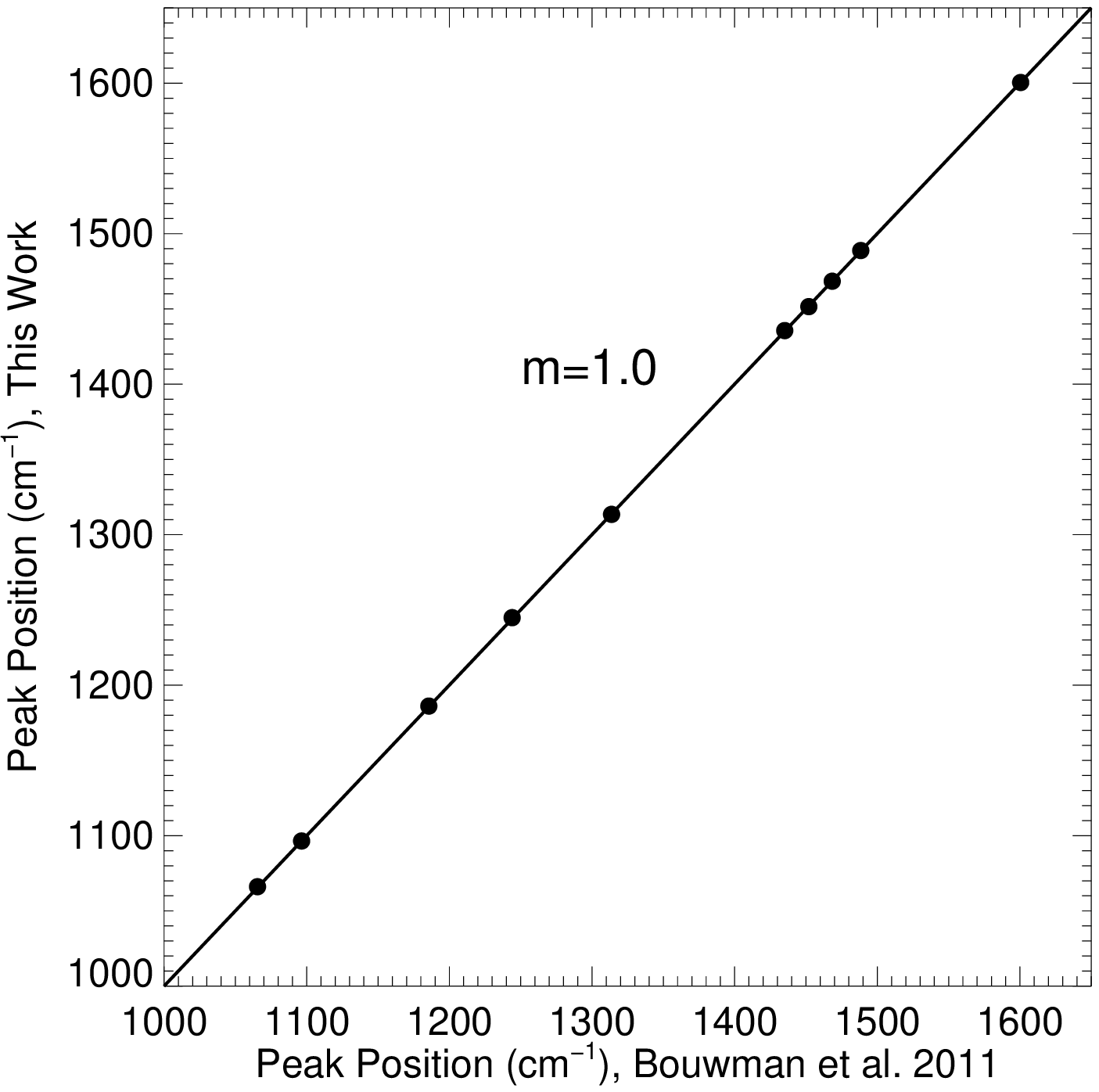} & \includegraphics[angle=0,scale=.45]{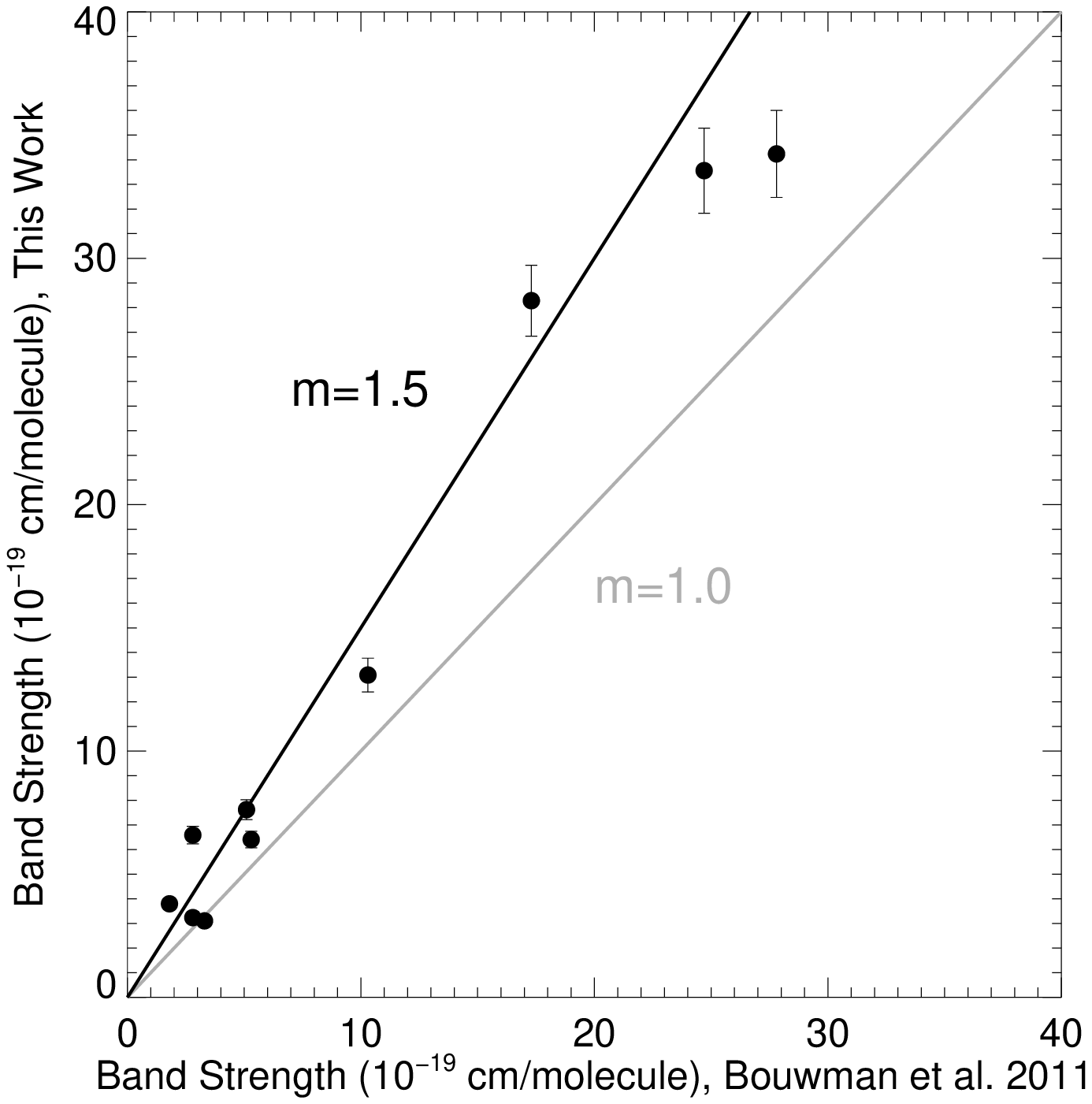} \\
	\end{tabular}
	\caption{Peak position (left panel) and band strength (right panel) comparison between this work and that of \citet{2011A&A...525A..93B}. 
	The error bars along the $y$-axis in the left panel are smaller than the plot points. Our calculated band strengths are 
	roughly 1.5 times those published by \citet{2011A&A...525A..93B}.}
	\label{fig:boVus}
\end{figure}

Figure \ref{fig:boVus} compares our results with those published by \citet{2011A&A...525A..93B}. 
While the positions of features measured in both works are essentially identical, the band strengths we report 
are $\sim$50\% larger than those reported by \citet{2011A&A...525A..93B}. Apparently, the theoretical band 
strength calculations used by \citet{2011A&A...525A..93B} consistently underestimate the values measured by 
our UV/IR correlation method.

%%%%%%%%%%%%%%%%%%%%%%%%%%%%%%%%%%%%%%%%%%%%%%%%%%%%%%%%%%%%%%%%%%%%%%%%%%%%

\section{Astrophysical Implications: PAH Absorption Features in YSO Spectra and the Carbon Budget}
\label{sec:ysos}

\begin{figure}[H]
	\centering
	\begin{tabular}{cc}
		\includegraphics[angle=0,scale=.45]{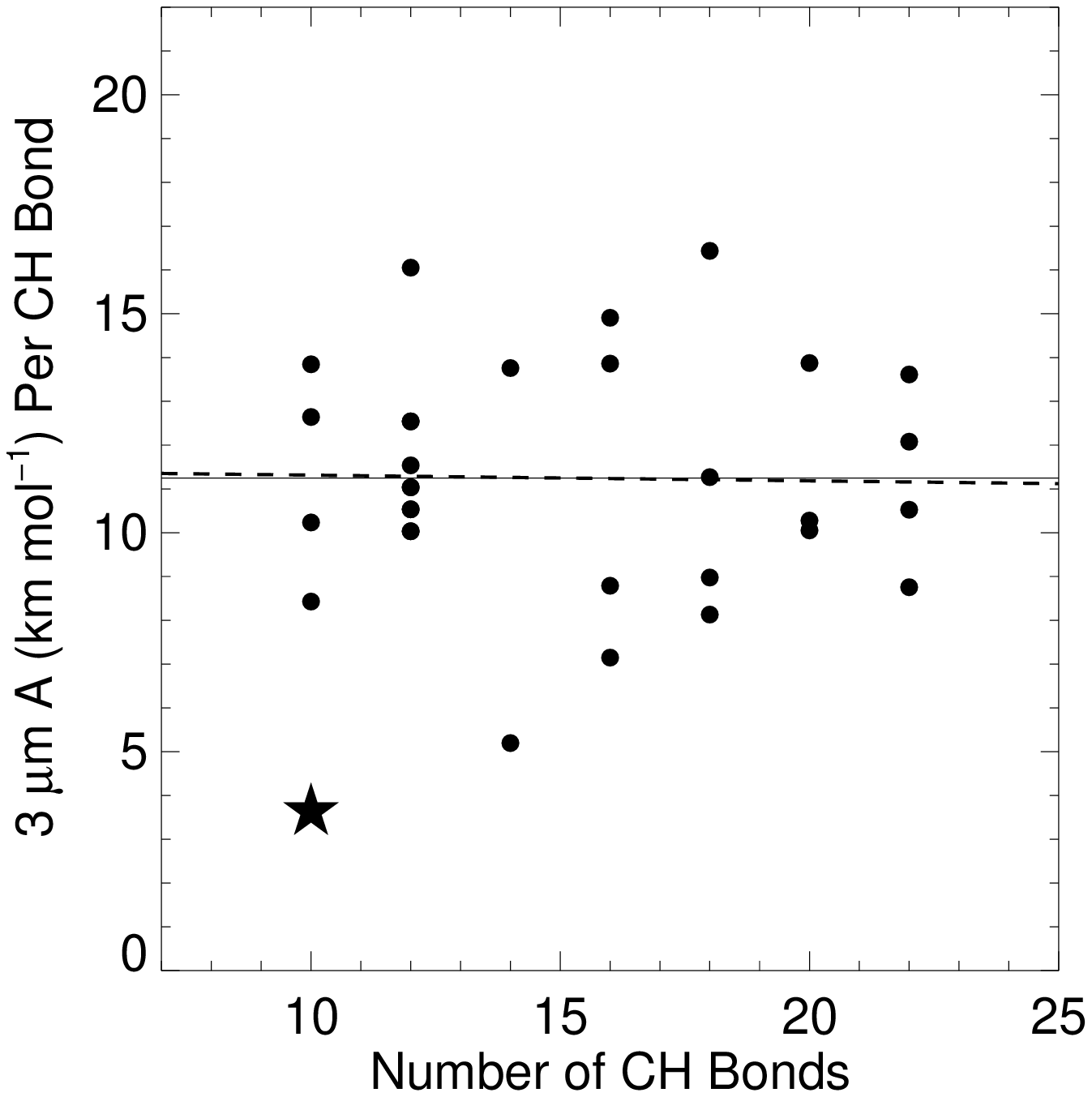} & \includegraphics[angle=0,scale=.45]{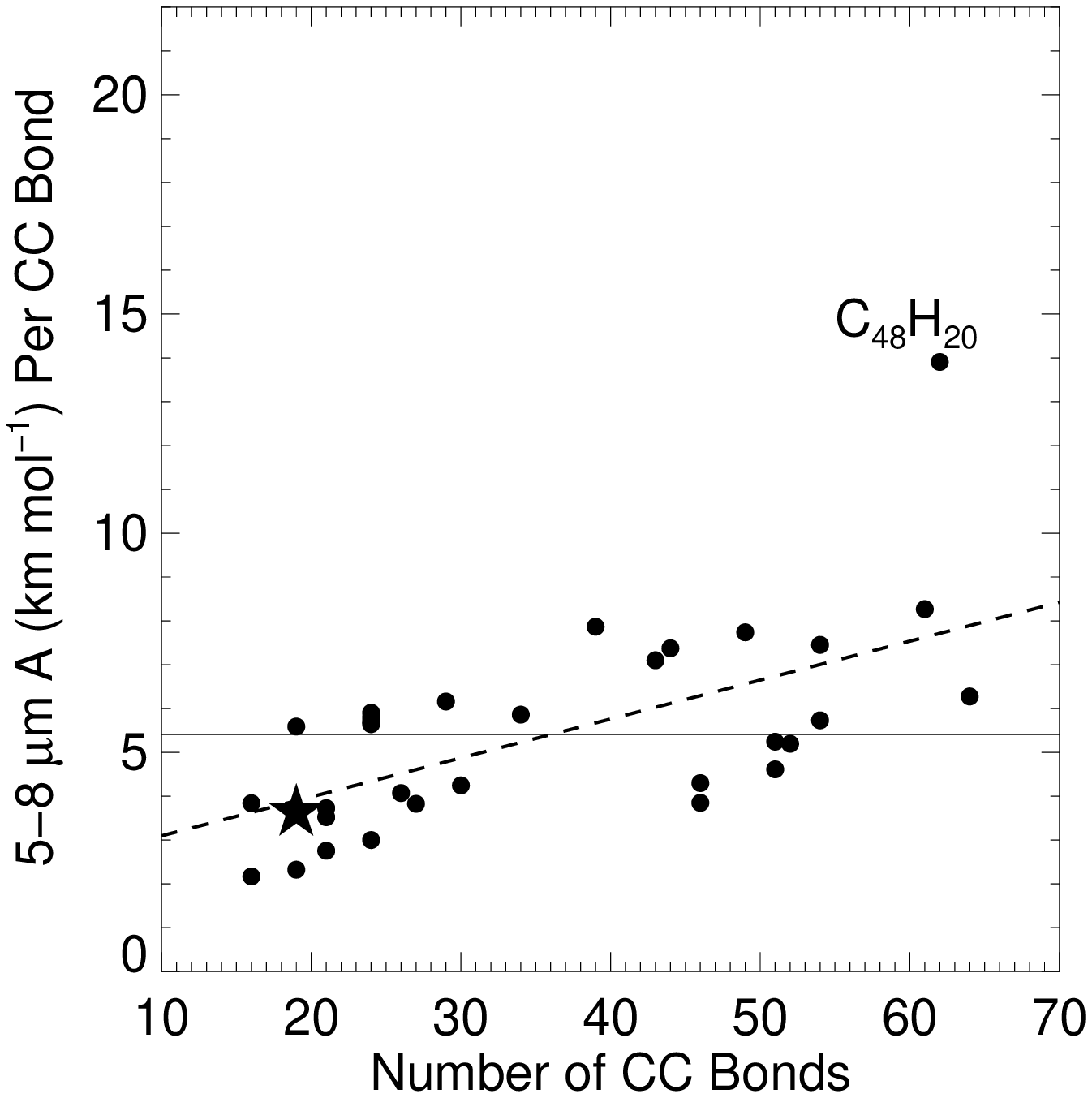} \\
	\end{tabular}
	\caption{Experimentally measured band strengths for the CH stretching mode near 3.25~\microns\ (left) and the CC stretching/CH in plane bending modes between 5$-$8~\microns\ (right) from the www.astrochem.org database \citep{2010ApJS..189..341B,2014BoersmaArticle} for neutral PAH species in argon (circles). Band strengths \textit{per bond} remain fairly constant across reported PAH species, although the 5$-$8~\microns\ band strengths do change mildly with a slope of 0.09~km~mol$^{-1}$ per CC bond per total number of CC bonds (excluding the outlier due to C$_{48}$H$_{20}$). Dashed lines show the best fitting correlations between band strengths per bond and number of bonds, while solid lines show the best fitting constant band strengths per bond. Stars mark the band strengths per bond for pyrene in water ice.}
	\label{fig:perbond}
\end{figure}

As illustrated here for pyrene, PAH IR spectra are characterized by a CH stretching feature 
near 3.25~\microns, CC stretching and CH in plane bending modes between 5 and 10~\microns, 
and CH out of plane bending modes between roughly 10 and 15~\microns. The detection of a 
3.25~\microns\ band, attributed to the CH stretching mode of PAHs, was reported for a 
handful of YSO spectra \citep{1994ApJ...433..179S,1996ApJ...459..209B,1999ApJ...517..883B}.
We looked for the presence of such a feature in the YSO spectra published by 
\citet{2008ApJ...678..985B}. For those spectra where a feature 
was detected, we attempted to quantify the PAH column density and the contribution of absorption by 
PAHs to the 5$-$8~\microns\ region. Since we only have laboratory data for pyrene, the following 
measurements are reported as a function of the number of CH and CC bonds present. This of course 
requires the assumption that band strengths \textit{per bond} remain relatively constant among different 
PAH species. Figure \ref{fig:perbond} shows the validity of this assumption. Here, experimentally determined band strengths \textit{per bond} for neutral PAHs in argon from the www.astrochem.org database are shown as a function of the number of 
CH and CC bonds per molecule. For the CH stretching and CC stretching/CH in plane bending modes, band strengths were integrated from 
3200$-$2800~\wavenum\ and 2000$-$1250~\wavenum, respectively. 

\subsection{3.25 \microns\ Feature}\label{sec:3micron}

We analyzed the 3.25~\microns\ absorption band in the YSO spectra published 
by \citet{2008ApJ...678..985B}. In summary, a 
total of 21 objects were observed with sufficient wavelength coverage to examine their spectra 
for a feature near 3.25~\microns. Six of those objects were massive YSOs observed with the Short Wavelength 
Spectrometer on board the \textit{Infrared Space Observatory (ISO)} \citep[also published by][]{2004ApJS..151...35G}. 
The others were low mass YSOs observed in the $L$-band with ground-based instruments and longward of 5~\microns\ 
with the \textit{Spitzer Space Telescope}.

Ground-based spectra were much noisier than the \textit{ISO} spectra, so they were first smoothed with the boxcar 
technique using a window of five points. Only a subset of the \textit{ISO} spectra analyzed included flux density error 
bars, which were subsequently propagated to determine error bars for the optical depth spectra. To 
determine error bars for the remaining optical depth spectra, a line was fit to a region $\pm$0.025~\microns\ on 
either side of each data point, and the standard deviation of that data set with respect to the linear fit was 
assigned as the error bar value for the relevant point.

For each spectrum, a low-order (third to fifth) polynomial was fit to a continuum region, 
defined in Table \ref{tab:continuum}, depending 
on the absorption features of each spectrum. Next, the residual spectra were fit with up to five Gaussians, 
located at $\sim$3.25, 3.32, 3.36, 3.47, and 3.53~\microns, using MPFIT and 
MPFITFUN \citep{2009ASPC..411..251M}. The 
3.47~\microns\ feature, often seen in YSO spectra, has been attributed to solo hydrogens attached to 
``diamond"-like carbon clusters \citep{1992ApJ...399..134A} or to ammonia hydrate \citep{2001A&A...365..144D}. 
The remaining model features at 3.32, 3.36, and 3.53~\microns\ correspond to the expected locations of solid 
CH$_{4}$, C$_2$H$_6$, and CH$_3$OH, respectively \citep[e.g.,][]{2003dge..conf.....W}. In the ground-based 
spectra, the 3.32~\microns\ feature is heavily contaminated by telluric CH$_4$ lines.

\begin{figure}[H]
	\centering
	\includegraphics[angle=0,scale=.5]{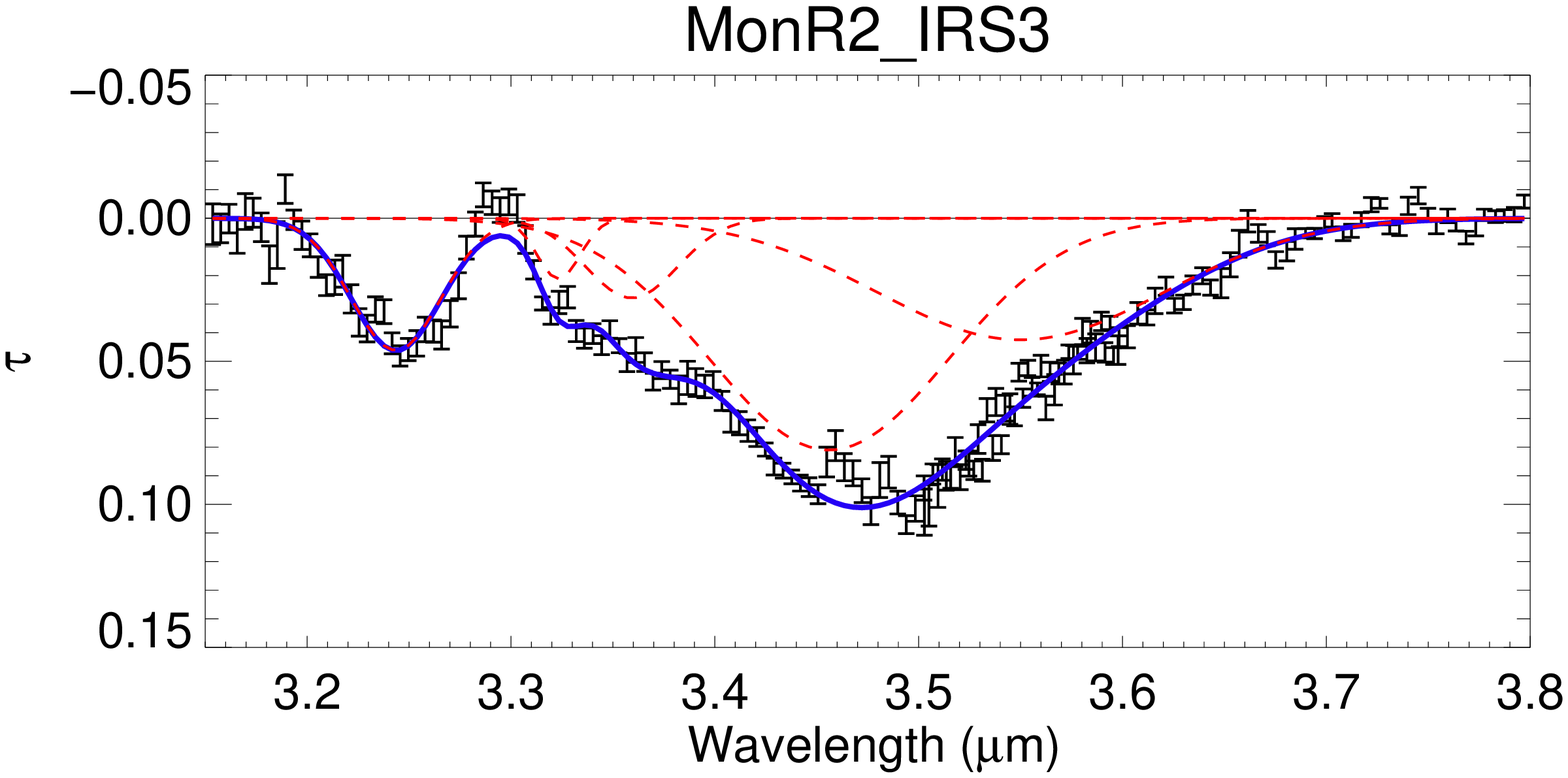}\\
	\includegraphics[angle=0,scale=.5]{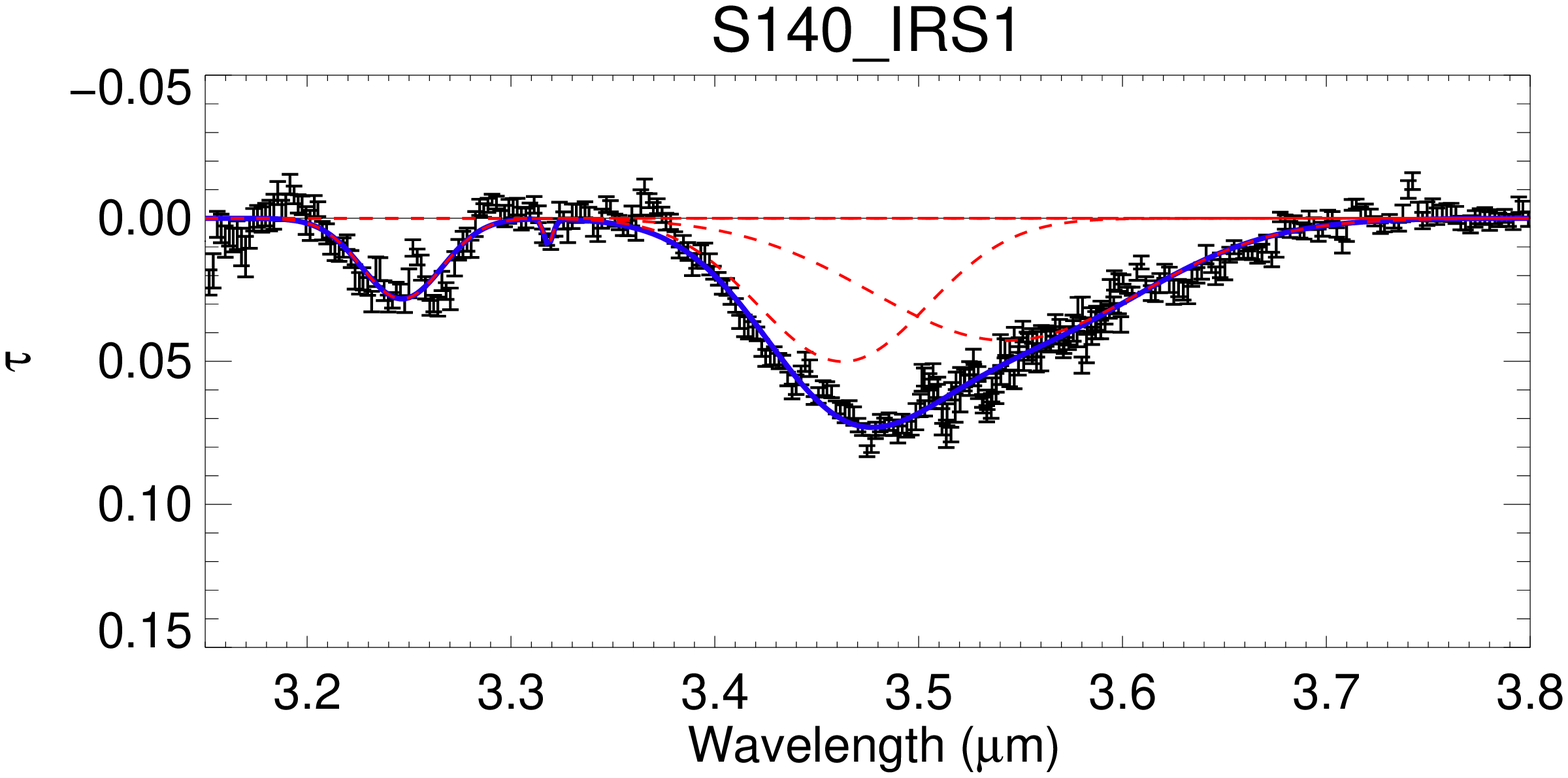}\\
	\includegraphics[angle=0,scale=.5]{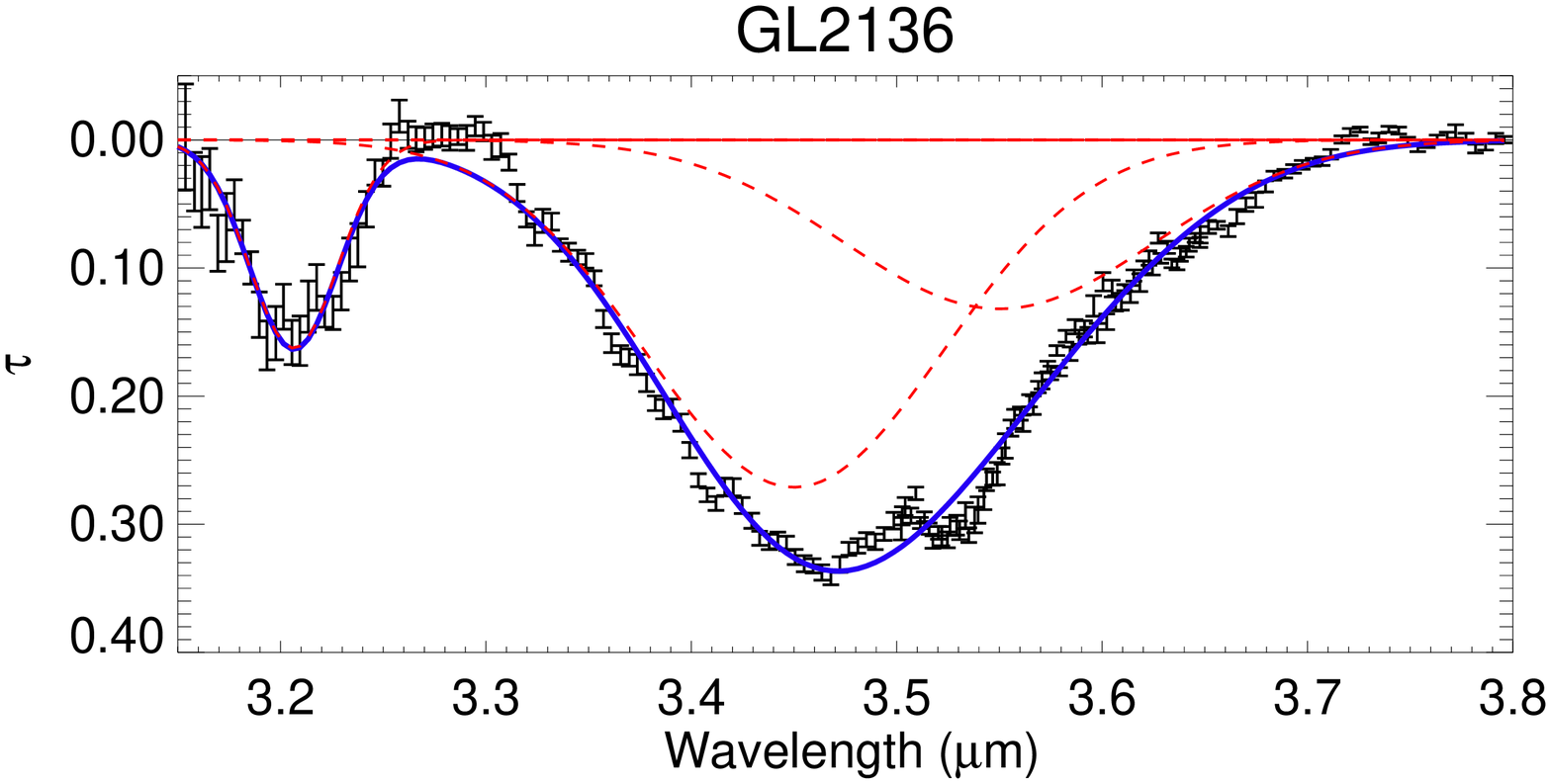}\\
	\caption{Optical depth spectral models after continuum subtraction. The model fits, composed of up to 5 Gaussians, are shown with solid blue lines. Each component of the model is shown separately with dashed red lines.}
	\label{fig:yso}
\end{figure}

Of the \textit{ISO} spectra, only three objects have a significant ($>3\sigma$) feature near 3.25~\microns. 
\citet{1999ApJ...517..883B}, using ground-based spectra, listed previously reported measurements 
of the feature for Mon~R2~IRS~3 and S140~IRS~1, 
but they excluded measurement of an apparent feature at 3.2~\microns\ for 
GL~2136 as they claim that its coincidence with the inflection point of crystalline ice absorption 
interferes with accurate measurement. These objects are shown in Figure \ref{fig:yso}, and the 
parameters of their model fits are listed in Table \ref{tab:yso}. The integrated optical depth 
measurement of the 3.25~\microns\ feature for Mon~R2~IRS~3 is in good agreement 
with previous results by \citet{1994ApJ...433..179S,1995ApJ...449L..69S}, and our measurement 
for S140~IRS~1 is similar to that reported by \citet{1996ApJ...459..209B} to within a factor of two.

\begin{figure}[H]
	\centering
	\includegraphics[angle=0,scale=.6]{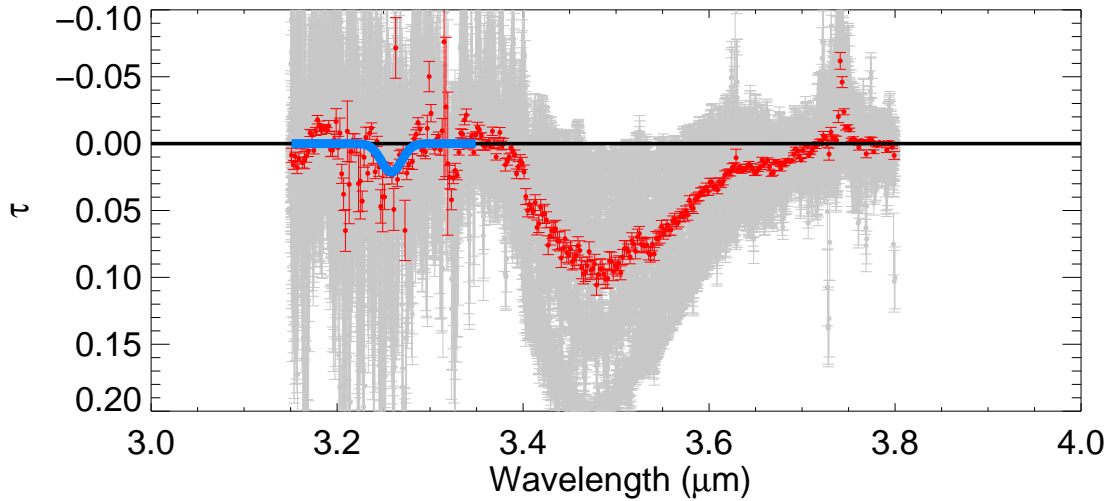}
	\caption{Average residual spectrum from ground-based observations, binned to 0.002~\microns\ (red). The 
	individual residual spectra after continuum subtraction are all overplotted in gray. A Gaussian, fit to the 
	prospective PAH feature, is shown in blue. The sharp feature near 3.32~\microns\ is due to telluric CH$_4$.}
	\label{fig:avgyso}
\end{figure}

For our ground-based YSO spectra, in order to increase the signal-to-noise ratio, all spectra were 
averaged to recover a global 3.25~\microns\ feature. The average, continuum subtracted 
spectrum, binned to 0.002~\microns, is shown in Figure \ref{fig:avgyso} on an optical depth scale. 
The best fit Gaussian to the region of interest is located at $3.258\pm0.004$~\microns\ with a maximum optical 
depth of $0.021\pm0.006$ and a FWHM of $24.13\pm6.93$~cm$^{-1}$ ($0.026\pm0.007$~\microns).

\clearpage

\begin{deluxetable}{ccccccc}
\tabletypesize{\small}
\tablecolumns{7}
\tablewidth{0pt}
\tablecaption{Continuum Definition
	\label{tab:continuum}}
\tablehead{
\colhead{Object} & \colhead{R.A.\ and Decl.} & \colhead{} & \colhead{} & \multicolumn{3}{c}{Continuum Regions} \\
\colhead{} 	    & \colhead{(J2000)}        & \colhead{} & \colhead{} & \multicolumn{3}{c}{(\microns)}}
\startdata
	Mon R2 IRS 3 & 06\hour07\min47\fs80 $-$06\deg22\arcmin55\farcs0 & & & 3.15$-$3.20 & 3.28$-$3.30 & 3.70$-$3.80 \\
	S140 IRS 1     & 22\hour19\min18\fs17 $+$63\deg18\arcmin47\farcs6 & & & 3.15$-$3.20 & 3.28$-$3.36 & 3.70$-$3.80 \\
	GL 2136        & 18\hour22\min26\fs32 $-$13\deg30\arcmin08\farcs2 & & & 3.14$-$3.17 & 3.25$-$3.33 & 3.70$-$3.80 \\
	Ground-based spectra & . . .                        & & & 3.15$-$3.20 & 3.28$-$3.36 & 3.70$-$3.80 \\
\enddata
\end{deluxetable}

\begin{deluxetable}{ccccc}
\tabletypesize{\small}
\tablecolumns{5}
\tablewidth{0pt}
\tablecaption{3.25~\microns\ Absorption Feature
	\label{tab:yso}}
\tablehead{
   \colhead{Object} & \colhead{$\lambda$}             & \multicolumn{2}{c}{FWHM}                            & \colhead{$\tau_{\rm max}$}\\
   \colhead{}           & \colhead{\microns}  & \colhead{\microns} & \colhead{cm$^{-1}$} & \colhead{}}
   \startdata
	Mon R2 IRS 3               & 3.244 $\pm$ 0.002 & 0.052 $\pm$ 0.004 & 49.3 $\pm$ 3.4 & 0.046 $\pm$ 0.003\\
	S140 IRS 1                    & 3.247 $\pm$ 0.001 & 0.046 $\pm$ 0.003 & 44.1 $\pm$ 3.1 & 0.028 $\pm$ 0.002\\
	GL 2136                       & 3.206 $\pm$ 0.003 & 0.051 $\pm$ 0.006 & 49.4 $\pm$ 6.0 & 0.163 $\pm$ 0.018\\
	Ground-based average & 3.258  $\pm$ 0.004 & 0.026 $\pm$ 0.007 & 24.1 $\pm$ 6.9 & 0.021 $\pm$ 0.006\\
   \enddata
\end{deluxetable}

\subsection{5$-$8~\microns\ Region}

\citet{2008ApJ...678..985B} found that the 5$-$8~\microns\ absorption regions in YSO spectra are 
due to at least five distinct components in addition to the bending mode of \h2o\ ice, each of which 
likely have multiple carriers. Since PAHs exhibit a number of 
features in this same region due to CC stretching and CH in plane bending modes, we 
hypothesized that some of the absorption 
could be due to PAHs frozen in ice mantles. 

\begin{figure}[H] 
	\centering 
	\includegraphics[angle=0,scale=.7]{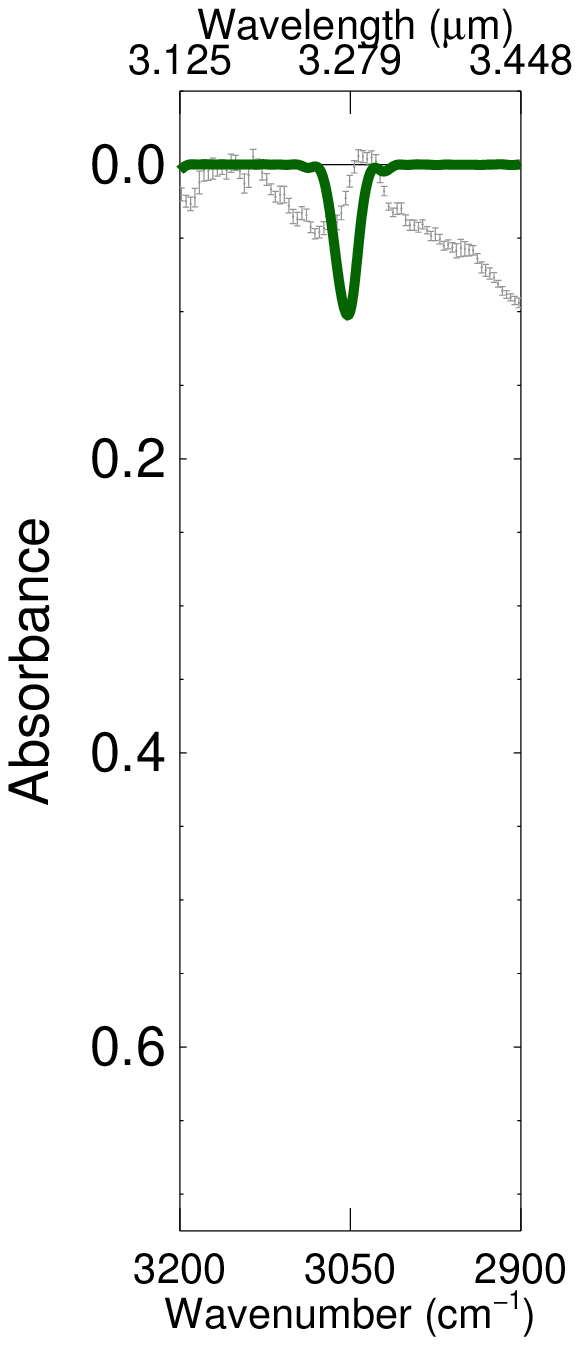} 
	\includegraphics[angle=0,scale=.7]{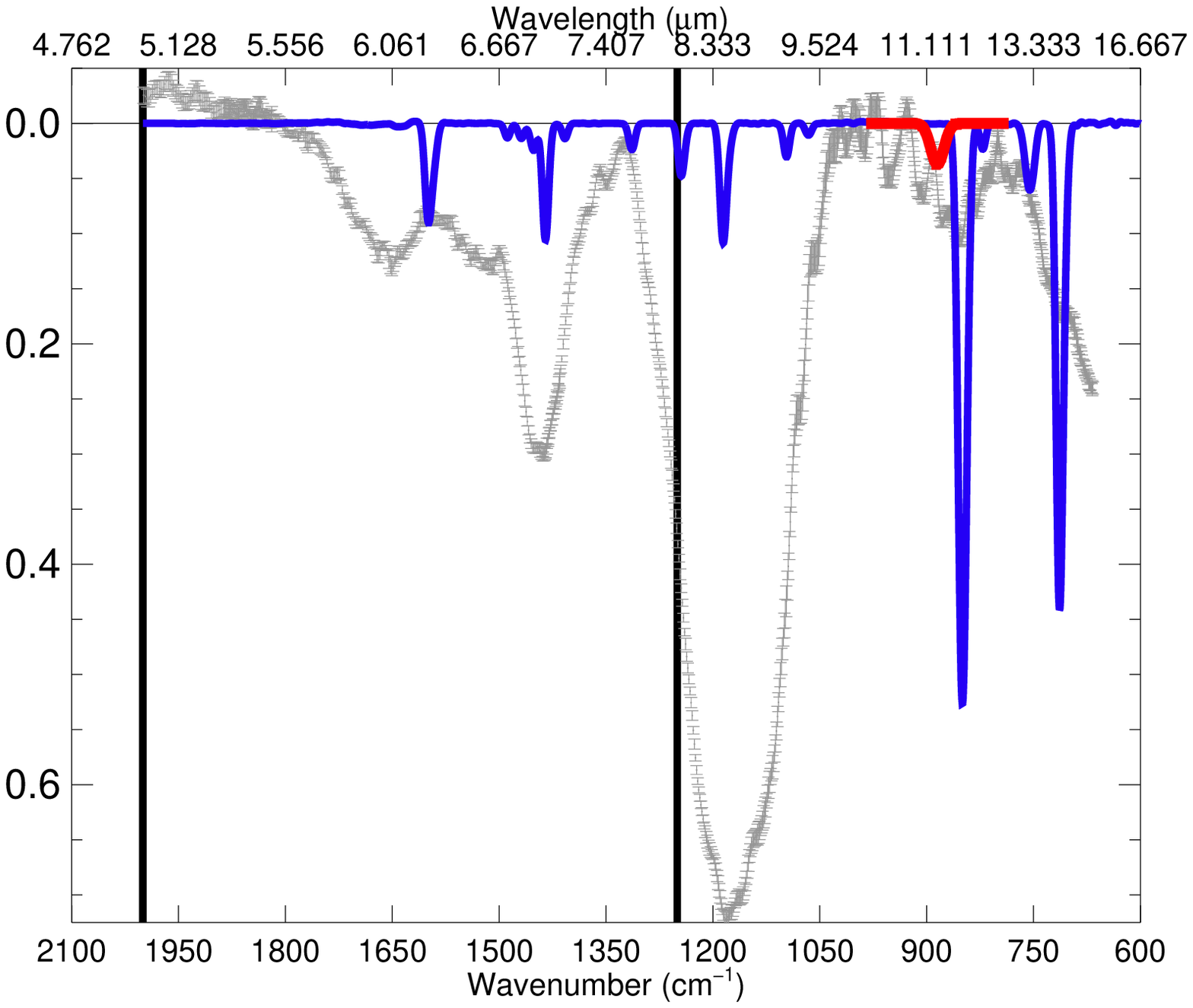} 
	\caption{Observed absorption spectrum (gray) of Mon R2 IRS 3 is compared to the laboratory 
spectrum of pyrene. The left panel shows the residual 3.25~\microns\ 
feature after subtraction of a local baseline (see Section~\ref{sec:3micron}). The YSO spectrum from 5 to 15~\microns\ 
is plotted after removal of \h2o\ ice and silicate absorption \citep{2008ApJ...678..985B}. The Gaussian at 
11.3~\microns\ (red) corresponds to the PAH detection reported by \citet{2000ApJ...544L..75B}. The pyrene 
spectrum (green/blue) was scaled so that the area of band A is equivalent to the area of the corresponding YSO feature. 
All observational and experimental spectra were convolved to a matching resolution of $\sim$8~\wavenum. 
Vertical lines denote the 5$-$8~\micron\ region where we estimated the PAH contribution to 
observed absorption.}
	\label{fig:lab} 
\end{figure}

Assuming that any 3.25~\microns\ feature present is due to neutral PAHs frozen in ice, 
we used our previous calculations of absolute band strengths for pyrene to constrain the potential 
PAH contribution to the 5$-$8~\microns\ absorption region ($f_{\rm PAH}$, see Table~\ref{tab:percent}), 
after subtraction of the \h2o\ bending mode. Ours is only a first-order 
approximation since the PAHs present in the ISM are likely to consist of a mixture of many species, as 
clearly illustrated by Figure~\ref{fig:lab}. 

The average PAH molecule in the ISM contains 50 carbon atoms 
\citep{2008ARA&A..46..289T}, rather than the 16 of pyrene. To account for this we assume that 
the band strengths \textit{per bond} remain constant amongst different PAH species, although computational spectra 
predict the CH stretching band to increase for large PAHs \citep{2008ApJ...678..316B}, and we report 
contribution percentages in terms of the ratio of CC to CH bonds which characterizes the actual mixture 
of observed interstellar PAHs. For reference, that ratio is $\frac{n_{\rm{CC}}}{n_{\rm{CH}}}=1.9$ for 
pyrene. The simplest PAH, naphthalene (C$_{10}$H$_{8}$), has a ratio of 
$\frac{n_{\rm{CC}}}{n_{\rm{CH}}}=1.375$, and circumcoronene (C$_{54}$H$_{18}$), the most compact 
and radially symmetric PAH with $\sim$50 carbon atoms, has a ratio 
of $\frac{n_{\rm{CC}}}{n_{\rm{CH}}}=4$. Equations (\ref{eq:5to8a})$-$(\ref{eq:5to8c}) spell out how to calculate 
$f_{\rm PAH}$. In these equations, $\Delta x$ refers to the FWHM 
of a particular feature. Band strengths from bands B$-$H in 
Table~\ref{tab:final} were summed and divided by 19 (the number of CC bonds in pyrene) to obtain a 
band strength of $A_{\rm{CC}}=4.497\ (\pm0.234) \times 10^{-19}$~cm \textit{per CC bond}. 
The band strength of band A is $A_{\rm{CH}}=6.036\ (\pm0.312) \times 10^{-19}$~cm \textit{per CH bond}.

\begin{equation}
N({\rm PAH})=\frac{\tau\Delta x_{3\tiny{\microns}}}{A_{\rm{CH}}n_{\rm{CH}}}=\frac{\Sigma_{5-8\tiny{\microns}}\tau\Delta x}{A_{\rm{CC}}n_{\rm{CC}}}
\label{eq:5to8a}
\end{equation}

\begin{equation}
\Sigma_{5-8\tiny{\microns}}\tau\Delta x=\tau\Delta x_{3\tiny{\microns}}\frac{A_{\rm{CC}}}{A_{\rm{CH}}}\frac{n_{\rm{CC}}}{n_{\rm{CH}}}
\label{eq:5to8b}
\end{equation}

\begin{equation}
f_{\rm PAH}=\frac{\Sigma_{5-8\tiny{\microns}}\tau\Delta x}{\int{\tau dx}_{\rm obs}}\times100
\label{eq:5to8c}
\end{equation}

\begin{deluxetable}{cccccc}
\tabletypesize{\small}
\tablecolumns{6}
\tablewidth{0pt}
\tablecaption{Percent Contribution to the 5$-$8~\microns\ Region ($f_{\rm PAH}$) After \h2o\ Subtraction
	\label{tab:percent}}
\tablehead{
\colhead{Object} & \colhead{$N$(CH) (10$^{18}$~cm$^{-2}$)} & \multicolumn{2}{c}{$N$(CC) (10$^{18}$~cm$^{-2}$)}                                        & \multicolumn{2}{c}{$f_{\rm PAH}$ (\%)}\\
\colhead{} & \colhead{} & \colhead{$\frac{n_{\rm{CC}}}{n_{\rm{CH}}}=1.375$} & \colhead{$\frac{n_{\rm{CC}}}{n_{\rm{CH}}}=4$} & \colhead{$\frac{n_{\rm{CC}}}{n_{\rm{CH}}}=1.375$} & \colhead{$\frac{n_{\rm{CC}}}{n_{\rm{CH}}}=4$}}
\startdata
	Mon R2 IRS 3                         & 3.76 $\pm$ 0.41  & 5.17 $\pm$ 0.56   & 15.03 $\pm$ 1.62  & 2.9 & 8.5 \\
	S140 IRS 1                             & 2.05 $\pm$ 0.23  & 2.81 $\pm$ 0.32    & 8.18 $\pm$ 0.92    & 2.0 & 6.0 \\
	GL 2136$^{\rm{\textbf{a}}}$ & 13.34 $\pm$ 2.30 & 18.34 $\pm$ 3.16 & 53.36 $\pm$ 9.18 & 6.9 & 20.4 \\
\enddata
\vspace{-0.8cm}
\tablecomments{$^{\rm{\textbf{a}}}$\citet{1999ApJ...517..883B} report that the inflection point of a crystalline ice absorption profile coincides with the apparent ``feature'' in GL~2136, so the calculated abundances reported here for this system may be inaccurate.}
%%% bocomp.pro makes these calculations!!!

\end{deluxetable}

\subsection{Carbon Budget}
\label{sec:carbon}

\begin{figure}[H]
	\centering
	\includegraphics[angle=0,scale=.55]{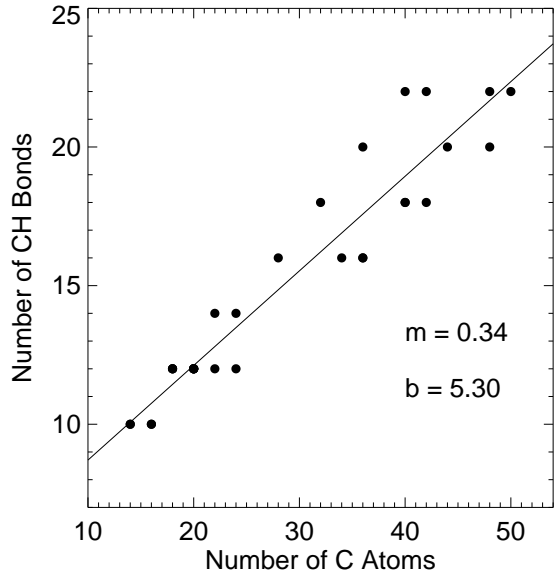} 
	\caption{Relationship between number of CH bonds and number of carbon atoms for PAHs ranging in size from C$_{14}$H$_{10}$ through C$_{50}$H$_{22}$, as listed in the experimental section on www.astrochem.org \citep{2010ApJS..189..341B,2014BoersmaArticle}. According to this relationship, an average-sized PAH containing 50 C atoms \citep{2008ARA&A..46..289T} should have roughly 22 CH bonds.}
	\label{fig:CHvsC}
\end{figure}

Observations of PAH emission from the diffuse ISM indicate that 10\%$-$20\% of the total carbon 
is locked in PAH molecules \citep{2011IAUS..280..149P}. The column densities of CH bonds 
reported in Table~\ref{tab:percent} can be used to calculate the fraction of the carbon abundance 
locked up in PAHs in the ices around YSOs. Hydrogen column densities reported by \citet{2013ApJ...777...73B} were converted 
to carbon column densities using $x(\rm{C})=3.9\times10^{-4}$ \citep{2008ARA&A..46..289T}. 
The number of carbon atoms locked in PAHs were calculated from $N$(CH) assuming that the 
average PAH molecule contains $\sim$22 CH bonds. That value was determined using 
Figure~\ref{fig:CHvsC} assuming the average PAH contains 50 carbon atoms \citep{2008ARA&A..46..289T}. 
Table~\ref{tab:carbon} shows the estimates for the percentages of carbon locked in neutral PAHs around YSOs. 
\citet{1995ApJ...449L..69S} estimated this value for Mon~R2~IRS~3 to be about 4\% which is half of the value 
reported in this work, but they also used a higher band strength for the CH bond of 
$1.7\times10^{-18}$~cm/bond. With a larger sample of YSOs, \citet{1999ApJ...517..883B} estimate the 
percentage of carbon in PAHs to range from 8\%$-$23\%. 

We chose the CH stretching bands to constrain PAH abundances because they tend to be most consistent across PAH species, but \citet{2001A&A...376..254K} estimated PAH abundances toward YSOs using excess absorption at 6.2~\microns\ instead. Their estimates for the percentage of total carbon locked in PAHs ranged from 83\%, assuming neutral PAHs, to 7\%, assuming ionized PAHs. This vast difference is due to the fact that neutral PAHs have relatively strong 3.25~\microns\ features and weak 5$-$10~\microns\ features, while ionized PAHs exhibit the opposite behavior. According to our estimates, Mon~R2~IRS~3 and S140~IRS~1 have less than 10\% of their total carbon in neutral PAHs. These estimates are upper limits in terms of neutral PAH abundances since they assumed the entire 3.25~\microns\ features were due solely to neutral PAHs. 
If the environments near YSOs follow the same trends as the rest of the galaxy, namely that PAHs in these systems contain 10\%$-$20\% of the carbon reservoir, ionized PAHs should account for the 
remaining PAH contribution to the carbon budget. When ionized PAHs are accounted for, the portion of 
absorption in the 5$-$8~\microns\ spectral region due to PAHs should be even higher than what is 
reported in Table~\ref{tab:percent}. The estimate for GL~2136 that neutral PAHs account for 22\% of its total 
carbon seems high, lending support to the 
conclusion of \citet{1999ApJ...517..883B} that absorption from crystalline \h2o\ ice is responsible for the 
apparent 3.2~\microns\ ``feature'' in its spectrum. Therefore, caution must be exercised when interpreting 
the cause of absorption in this region.

\begin{deluxetable}{cccccc}
\tabletypesize{\small}
\tablecolumns{6}
\tablewidth{0pt}
\tablecaption{Carbon Budget for Neutral PAHs Around YSOs
	\label{tab:carbon}}
\tablehead{
\colhead{Object} & \colhead{$N_{\rm H}^{\rm{\textbf{a}}}$} & \colhead{$N_{\rm C}$, total} & \colhead{$N$(CH)} & \colhead{$N_{\rm C}$, in PAHs} & \colhead{\% Carbon} \\
\colhead{} & \colhead{(10$^{22}$~cm$^{-2}$)} & \colhead{(10$^{18}$~cm$^{-2}$)} & \colhead{(10$^{18}$~cm$^{-2}$)} & \colhead{(10$^{18}$~cm$^{-2}$)} & \colhead{in Neutral PAHs}}
\startdata
	Mon R2 IRS 3 & 26.19 & 102.14 & 3.76   & 8.55   & 8.37\\
	S140 IRS 1     & 22.41 & 87.40   & 2.05   & 4.66   & 5.33\\
	GL 2136$^{\rm{\textbf{b}}}$        & 35.20 & 137.67  & 13.34 & 30.32 & 22.02\\
\enddata
\vspace{-0.8cm}
\tablecomments{$^{\rm{\textbf{a}}}$Reported by \citet{2013ApJ...777...73B}. $^{\rm{\textbf{b}}}$\citet{1999ApJ...517..883B} report that the inflection point of a crystalline ice absorption profile coincides with the apparent ``feature'' in GL~2136, so the calculated abundances reported here for this system may be inaccurate.}
\end{deluxetable}
%%% carbon_estimate.pro makes these calculations!!!

%%%%%%%%%%%%%%%%%%%%%%%%%%%%%%%%%%%%%%%%%%%%%%%%%%%%%%%%%%%%%%%%%%%%%%%%%%%%

\section{Summary and Future Work}
\label{sec:summary}

PAHs account for 10\%$-$20\% of the Milky Way's carbon reservoir, yet their direct observation via 
absorption in the dense ISM remains difficult. Quantification of PAHs expected to be 
embedded in interstellar ices is hindered by the limited knowledge of the absolute band strengths, 
peak wavelengths, and band widths for PAHs frozen in ice matrices. Our work addresses this issue via 
an investigation of pyrene embedded in water ice.
\begin{enumerate}
\item We report the first laboratory determination of the band strength for the CH stretching mode 
of pyrene in water ice near 3.25~\microns.
\item We report new absolute band strengths from 3$-$15~\microns\ for pyrene embedded in \h2o\ 
and \d2o\ ice. The band strengths reported here are roughly 50\% greater than those 
published by \citet{2011A&A...525A..93B}. 
\item We combed the data set of YSO spectra 
published by \citet{2008ApJ...678..985B} to look for evidence of PAH absorption at 
3.25~\microns\ and used the results from our laboratory measurements to estimate PAH column 
densities where applicable. In our sample, neutral PAHs account for 2\%$-$9\% of the non-\h2o\ 
ice absorption from 5$-$8~\microns. Neutral PAHs account for 5\%$-$9\% of 
the carbon budget toward YSOs, depending on the particular PAH mixture 
present. 
\end{enumerate}

Further laboratory work is needed to determine widths, peak positions, and band strengths for a large 
variety of PAH species embedded in ices. 
Figure~\ref{fig:lab} shows why this is critical. It is obvious from a comparison of the observed 
absorption of Mon R2 IRS 3 to our laboratory measurements that pyrene alone cannot explain the feature at 
3052~\wavenum\ (3.25~\microns) attributed to CH stretching. Band A in our laboratory spectra 
is too narrow and slightly offset from the observed 3.25~\microns\ feature. Beyond 10~\microns, 
the CH out of plane bending features for pyrene do not 
align with the PAH detection at 11.3~\microns\ reported by \citet{2000ApJ...544L..75B}. In addition, the 
pyrene features in this region are much stronger than the reported 11.3~\microns\ feature. This apparent discrepancy supports the conclusion that PAHs around Mon R2 IRS 3 must contain a mixture of species. As illustrated in Figure~2 of \citet{1999ApJ...511L.115A}, a mixture of neutral PAHs will exhibit a strong, narrow feature at 3.25~\microns\ and much broader, weaker features beyond 10~\microns. This is because the 
position of the CH stretching mode is quite stable across PAH species whereas the CH out of plane bending 
modes, though stronger for any individual species, are much more variable in position. The result is an 
apparent broadening and weakening of the composite absorption feature beyond 10~\microns\ relative to 
the 3.25~\microns\ feature in the spectrum of a PAH mixture.

Future work should also investigate the effects of other relevant ice matrices (e.g., CO$_2$) on the PAH 
absorption profiles. Eventually, the infrared band strengths for a variety of ionized PAHs need to be measured to compliment work with neutral species. Finally, the 
greatest difficulty for detecting PAH absorption features in YSO spectra is the lack of sufficient 
signal-to-noise due to atmospheric effects (in particular telluric CH$_4$ lines) in the ground-based data. Observational campaigns 
during the upcoming \textit{SOFIA} and \textit{James Webb Space Telescope} missions should effectively solve this problem.

%%%%%%%%%%%%%%%%%%%%%%%%%%%%%%%%%%%%%%%%%%%%%%%%%%%%%%%%%%%%%%%%%%%%%%%%%%%%

\section*{\textit{Acknowledgements}}

\textit{This research was carried out at the Jet Propulsion Laboratory and IPAC, California 
Institute of Technology, under a contract with the National Aeronautics and Space Administration. Support 
for this research was provided in part by funding from the NASA Astrobiology Institute to Rensselaer 
Polytechnic Institute (award NNA09DA80A). The experimental part of this work was enabled through 
partial funding from JPL’s DRDF and R\&TD 
funding for infrastructure of the ``Ice Spectroscopy Laboratory" at the Jet Propulsion Laboratory and was 
carried out by H.L.\ and M.S.G., supported by an astrophysics laboratory research award funded by the Spitzer Space Telescope. H.L.\ also acknowledges the 
Finnish Cultural Foundation for financial support. We thank 
Dr.\ Irene Li Barnett and Dr.\ Antti Lignell who were involved in building the experimental set-up used and
initial training of H.L.\ Data analysis was carried out by E.H-U.\ at California Institute of Technology 
(supervised by A.B.\ and M.S.G.) funded by an IPAC Visiting Graduate Student Fellowship.}

%%%%%%%%%%%%%%%%%%%%%%%%%%%%%%%%%%%%%%%%%%%%%%%%%%%%%%%%%%%%%%%%%%%%%%%%%%%%
%Create ApJ-type bibliography
\bibliographystyle{apj} 
%Use the master file of all references ever used 
\bibliography{MASTERrefs}

\begin{thebibliography}{42}
\expandafter\ifx\csname natexlab\endcsname\relax\def\natexlab#1{#1}\fi

\bibitem[{{Allamandola}(2011)}]{2011EAS....46..305A}
{Allamandola}, L.~J. 2011, in EAS Publications Series, Vol.~46, EAS
  Publications Series, ed. C.~{Joblin} \& A.~G.~G.~M. {Tielens}, 305

\bibitem[{{Allamandola} {et~al.}(1999){Allamandola}, {Hudgins}, \&
  {Sandford}}]{1999ApJ...511L.115A}
{Allamandola}, L.~J., {Hudgins}, D.~M., \& {Sandford}, S.~A. 1999, ApJL, 511,
  L115

\bibitem[{{Allamandola} {et~al.}(1992){Allamandola}, {Sandford}, {Tielens}, \&
  {Herbst}}]{1992ApJ...399..134A}
{Allamandola}, L.~J., {Sandford}, S.~A., {Tielens}, A.~G.~G.~M., \& {Herbst},
  T.~M. 1992, \apj, 399, 134

\bibitem[{{Barnett} {et~al.}(2012){Barnett}, {Lignell}, \&
  {Gudipati}}]{2012ApJ...747...13B}
{Barnett}, I.~L., {Lignell}, A., \& {Gudipati}, M.~S. 2012, \apj, 747, 13

\bibitem[{{Bauschlicher} {et~al.}(2008){Bauschlicher}, {Peeters}, \&
  {Allamandola}}]{2008ApJ...678..316B}
{Bauschlicher}, Jr., C.~W., {Peeters}, E., \& {Allamandola}, L.~J. 2008, \apj,
  678, 316

\bibitem[{{Bauschlicher} {et~al.}(2010){Bauschlicher}, {Boersma}, {Ricca},
  {Mattioda}, {Cami}, {Peeters}, {S{\'a}nchez de Armas}, {Puerta Saborido},
  {Hudgins}, \& {Allamandola}}]{2010ApJS..189..341B}
{Bauschlicher}, Jr., C.~W., {et~al.} 2010, \apjs, 189, 341

\bibitem[{{Berlman}(1971)}]{Berlman1971}
{Berlman}, I.~B. 1971, Handbook of Fluorescence Spectra of Aromatic Molecules
  (2nd ed.; New York: Academic Press)

\bibitem[{{Bernstein} {et~al.}(2002){Bernstein}, {Dworkin}, {Sandford},
  {Cooper}, \& {Allamandola}}]{2002Natur.416..401B}
{Bernstein}, M.~P., {Dworkin}, J.~P., {Sandford}, S.~A., {Cooper}, G.~W., \&
  {Allamandola}, L.~J. 2002, \nat, 416, 401

\bibitem[{{Bernstein} {et~al.}(2005){Bernstein}, {Sandford}, \&
  {Allamandola}}]{2005ApJS..161...53B}
{Bernstein}, M.~P., {Sandford}, S.~A., \& {Allamandola}, L.~J. 2005, \apjs,
  161, 53

\bibitem[{{Bernstein} {et~al.}(2007){Bernstein}, {Sandford}, {Mattioda}, \&
  {Allamandola}}]{2007ApJ...664.1264B}
{Bernstein}, M.~P., {Sandford}, S.~A., {Mattioda}, A.~L., \& {Allamandola},
  L.~J. 2007, \apj, 664, 1264

\bibitem[{{Boersma} {et~al.}(2014){Boersma}, {Bauschlicher}, {Ricca},
  {Mattioda}, {Cami}, {Peeters}, {S{\'a}nchez de Armas}, {Puerta Saborido},
  {Hudgins}, \& {Allamandola}}]{2014BoersmaArticle}
{Boersma}, C., {et~al.} 2014, ApJS, 211, in press

\bibitem[{{Boogert} {et~al.}(2013){Boogert}, {Chiar}, {Knez}, {{\"O}berg},
  {Mundy}, {Pendleton}, {Tielens}, \& {van Dishoeck}}]{2013ApJ...777...73B}
{Boogert}, A.~C.~A., {Chiar}, J.~E., {Knez}, C., {{\"O}berg}, K.~I., {Mundy},
  L.~G., {Pendleton}, Y.~J., {Tielens}, A.~G.~G.~M., \& {van Dishoeck}, E.~F.
  2013, \apj, 777, 73

\bibitem[{{Boogert} {et~al.}(2008){Boogert}, {Pontoppidan}, {Knez}, {Lahuis},
  {Kessler-Silacci}, {van Dishoeck}, {Blake}, {Augereau}, {Bisschop},
  {Bottinelli}, {Brooke}, {Brown}, {Crapsi}, {Evans}, {Fraser}, {Geers},
  {Huard}, {J{\o}rgensen}, {{\"O}berg}, {Allen}, {Harvey}, {Koerner}, {Mundy},
  {Padgett}, {Sargent}, \& {Stapelfeldt}}]{2008ApJ...678..985B}
{Boogert}, A.~C.~A., {et~al.} 2008, \apj, 678, 985

\bibitem[{{Bouwman} {et~al.}(2011){Bouwman}, {Mattioda}, {Linnartz}, \&
  {Allamandola}}]{2011A&A...525A..93B}
{Bouwman}, J., {Mattioda}, A.~L., {Linnartz}, H., \& {Allamandola}, L.~J. 2011,
  \aap, 525, A93

\bibitem[{{Bowman}(2006)}]{BowmanIDL}
{Bowman}, K.~P. 2006, An Introduction to Programming with IDL: Interactive Data
  Language (New York: Academic Press)

\bibitem[{{Bregman} {et~al.}(2000){Bregman}, {Hayward}, \&
  {Sloan}}]{2000ApJ...544L..75B}
{Bregman}, J.~D., {Hayward}, T.~L., \& {Sloan}, G.~C. 2000, ApJL, 544, L75

\bibitem[{{Brooke} {et~al.}(1999){Brooke}, {Sellgren}, \&
  {Geballe}}]{1999ApJ...517..883B}
{Brooke}, T.~Y., {Sellgren}, K., \& {Geballe}, T.~R. 1999, \apj, 517, 883

\bibitem[{{Brooke} {et~al.}(1996){Brooke}, {Sellgren}, \&
  {Smith}}]{1996ApJ...459..209B}
{Brooke}, T.~Y., {Sellgren}, K., \& {Smith}, R.~G. 1996, \apj, 459, 209

\bibitem[{{Dartois} \& {d'Hendecourt}(2001)}]{2001A&A...365..144D}
{Dartois}, E., \& {d'Hendecourt}, L. 2001, \aap, 365, 144

\bibitem[{{Dixon} {et~al.}(2005){Dixon}, {Taniguchi}, \& {Lindsey}}]{Dixon2005}
{Dixon}, J.~M., {Taniguchi}, M., \& {Lindsey}, J.~S. 2005, Photochemistry and
  Photobiology, 81, 212

\bibitem[{{Geers} {et~al.}(2009){Geers}, {van Dishoeck}, {Pontoppidan},
  {Lahuis}, {Crapsi}, {Dullemond}, \& {Blake}}]{2009A&A...495..837G}
{Geers}, V.~C., {van Dishoeck}, E.~F., {Pontoppidan}, K.~M., {Lahuis}, F.,
  {Crapsi}, A., {Dullemond}, C.~P., \& {Blake}, G.~A. 2009, \aap, 495, 837

\bibitem[{{Gibb} {et~al.}(2004){Gibb}, {Whittet}, {Boogert}, \&
  {Tielens}}]{2004ApJS..151...35G}
{Gibb}, E.~L., {Whittet}, D.~C.~B., {Boogert}, A.~C.~A., \& {Tielens},
  A.~G.~G.~M. 2004, \apjs, 151, 35

\bibitem[{{Gudipati}(2004)}]{2004JPC...108.4412G}
{Gudipati}, M.~S. 2004, JPCA, 108, 4412

\bibitem[{{Gudipati} \& {Allamandola}(2003)}]{2003ApJ...596L.195G}
{Gudipati}, M.~S., \& {Allamandola}, L.~J. 2003, ApJL, 596, L195

\bibitem[{{Gudipati} \& {Allamandola}(2004)}]{2004ApJ...615L.177G}
---. 2004, ApJL, 615, L177

\bibitem[{{Gudipati} \& {Allamandola}(2006)}]{2006ApJ...638..286G}
---. 2006, \apj, 638, 286

\bibitem[{{Gudipati} \& {Yang}(2012)}]{2012ApJ...756L..24G}
{Gudipati}, M.~S., \& {Yang}, R. 2012, ApJL, 756, L24

\bibitem[{{Kaiser} {et~al.}(2013){Kaiser}, {Stockton}, {Kim}, {Jensen}, \&
  {Mathies}}]{2013ApJ...765..111K}
{Kaiser}, R.~I., {Stockton}, A.~M., {Kim}, Y.~S., {Jensen}, E.~C., \&
  {Mathies}, R.~A. 2013, \apj, 765, 111

\bibitem[{{Keane} {et~al.}(2001){Keane}, {Tielens}, {Boogert}, {Schutte}, \&
  {Whittet}}]{2001A&A...376..254K}
{Keane}, J.~V., {Tielens}, A.~G.~G.~M., {Boogert}, A.~C.~A., {Schutte}, W.~A.,
  \& {Whittet}, D.~C.~B. 2001, \aap, 376, 254

\bibitem[{{Markwardt}(2009)}]{2009ASPC..411..251M}
{Markwardt}, C.~B. 2009, in Astronomical Society of the Pacific Conference
  Series, Vol. 411, Astronomical Data Analysis Software and Systems XVIII (San
  Francisco, CA: ASP)

\bibitem[{{Mu{\~n}oz Caro} {et~al.}(2002){Mu{\~n}oz Caro}, {Meierhenrich},
  {Schutte}, {Barbier}, {Arcones Segovia}, {Rosenbauer}, {Thiemann}, {Brack},
  \& {Greenberg}}]{2002Natur.416..403M}
{Mu{\~n}oz Caro}, G.~M., {et~al.} 2002, \nat, 416, 403

\bibitem[{{Peeters}(2011)}]{2011IAUS..280..149P}
{Peeters}, E. 2011, in IAU Symposium, Vol. 280, The Molecular Universe, ed.
  J.~{Cernicharo} \& R.~{Bachiller} (Cambridge: Cambridge Univ. Press), 149

\bibitem[{{Ray} {et~al.}(2006){Ray}, {Chakraborty}, \& {Moulick}}]{Rayetal2006}
{Ray}, G.~B., {Chakraborty}, I., \& {Moulick}, S.~P. 2006, Journal of Colloid
  and Interface Science, 294, 248

\bibitem[{{Salama} \& {Allamandola}(1992)}]{1992Natur.358...42S}
{Salama}, F., \& {Allamandola}, L.~J. 1992, \nat, 358, 42

\bibitem[{{Sandford} {et~al.}(2004){Sandford}, {Bernstein}, \&
  {Allamandola}}]{2004ApJ...607..346S}
{Sandford}, S.~A., {Bernstein}, M.~P., \& {Allamandola}, L.~J. 2004, \apj, 607,
  346

\bibitem[{{Sellgren} {et~al.}(1995){Sellgren}, {Brooke}, {Smith}, \&
  {Geballe}}]{1995ApJ...449L..69S}
{Sellgren}, K., {Brooke}, T.~Y., {Smith}, R.~G., \& {Geballe}, T.~R. 1995,
  ApJL, 449, L69

\bibitem[{{Sellgren} {et~al.}(1994){Sellgren}, {Smith}, \&
  {Brooke}}]{1994ApJ...433..179S}
{Sellgren}, K., {Smith}, R.~G., \& {Brooke}, T.~Y. 1994, \apj, 433, 179

\bibitem[{{Siu} \& {Duhamel}(2008)}]{SiuandDuhamel2008}
{Siu}, H., \& {Duhamel}, J. 2008, JPCB, 112, 15301

\bibitem[{{Thony} \& {Rossi}(1997)}]{1997PCPB..104..25T}
{Thony}, A., \& {Rossi}, M.~J. 1997, Journal of Photochemistry and Photobiology
  A-Chemistry, 104, 25

\bibitem[{{Tielens}(2008)}]{2008ARA&A..46..289T}
{Tielens}, A.~G.~G.~M. 2008, \araa, 46, 289

\bibitem[{{Whittet}(2003)}]{2003dge..conf.....W}
{Whittet}, D.~C.~B., ed. 2003, {Dust in the Galactic Environment (2nd ed.;
  Bristol: IOP)}

\bibitem[{{Williams} {et~al.}(2010){Williams}, {Bureau}, \&
  {Cappellari}}]{2010MNRAS.409.1330W}
{Williams}, M.~J., {Bureau}, M., \& {Cappellari}, M. 2010, \mnras, 409, 1330

\end{thebibliography}

\end{document}